\documentclass[a4paper,12pt]{article}
\pdfoutput=1
\usepackage{amsmath}
\usepackage{amssymb}
\usepackage{fullpage}
\usepackage{verbatim}

\usepackage[T1,OT1]{fontenc}

\usepackage{graphicx}
\usepackage{epstopdf}

\allowdisplaybreaks[1]

\usepackage[hang]{footmisc}
\usepackage[colorlinks,linktocpage,linkcolor=blue,citecolor=blue,urlcolor=blue]{hyperref}
\usepackage[compress,square,numbers]{natbib}

\newcommand{\e}{\mathrm{e}}

\begin{document}

\numberwithin{equation}{section}

\thispagestyle{empty}

\begin{flushright}
\small LMU-ASC 21/15\\
\normalsize
\end{flushright}
\vspace*{2cm}

\begin{center}

{\LARGE \bf Large-Field Inflation with Multiple Axions}

\vspace{0.5cm}

{\LARGE \bf and the Weak Gravity Conjecture}

\vspace{2cm}
{\large Daniel Junghans}\\

\vspace{1cm}
Arnold-Sommerfeld-Center f{\"{u}}r Theoretische Physik\\
Department f{\"{u}}r Physik, Ludwig-Maximilians-Universit{\"{a}}t M{\"{u}}nchen\\
Theresienstra\ss e 37, 80333 M{\"{u}}nchen, Germany

\vspace{1cm}
{\upshape\ttfamily daniel.junghans@lmu.de}\\

\vspace{1cm}
\begin{abstract}
In this note, we discuss the implications of the weak gravity conjecture (WGC) for general models of large-field inflation with a large number of axions $N$. We first show that, from the bottom-up perspective, such models admit a variety of different regimes for the enhancement of the effective axion decay constant, depending on the amount of alignment and the number of instanton terms that contribute to the scalar potential. This includes regimes of no enhancement, power-law enhancement and exponential enhancement with respect to $N$. As special cases, we recover the Pythagorean enhancement of $N$-flation, the $N$ and $N^{3/2}$ enhancements derived by Bachlechner, Long and McAllister and the exponential enhancement by Choi, Kim and Yun.
We then analyze which top-down constraints are put on such models from the requirement of consistency with quantum gravity. In particular, the WGC appears to imply that the enhancement of the effective axion decay constant must not grow parametrically with $N$ for $N \gg 1$. On the other hand, recent works proposed that axions might be able to violate this bound under certain circumstances. Our general expression for the enhancement allows us to translate this possibility into a condition on the number of instantons that couple to the axions. We argue that, at large $N$, models consistent with quantum gravity must either allow super-Planckian field excursions or have an enormous, possibly even exponentially large, number of dominant instanton terms in the scalar potential.
\end{abstract}

\end{center}

\newpage

\section{Introduction}

Even after the recent release of the combined Planck/BICEP2 results \cite{Ade:2015tva}, the updated bound $r < 0.12$ on the tensor-to-scalar ratio of primordial fluctuations is still well above the Lyth bound $r < 0.01$ \cite{Lyth:1996im} such that models of large-field inflation remain viable scenarios compatible with observations.
In such models, the inflaton undergoes a large field excursion in Planck units during inflation and, hence, operators of all mass dimensions are relevant in the inflaton potential. Models of large-field inflation are thus sensitive to quantum gravity effects, suggesting that they can consistently be studied only in a string theory framework (see \cite{Baumann:2014nda} for a review of inflation in string theory).

In this context, axions are particularly well-motivated inflaton candidates since they enjoy a continuous shift symmetry which protects the inflaton potential from perturbative quantum corrections \cite{Freese:1990rb}. The shift symmetry is broken non-perturbatively to a discrete subgroup such that the only contributions to the inflaton potential are \mbox{(multi-)}instanton terms. Moreover, axions arise naturally in string compactifications, e.g., from the dimensional reduction of Abelian $p$-form fields, where the shift symmetry is a 4D remnant of the higher-dimensional gauge symmetry.

Unfortunately though, it turns out that it is difficult to realize explicit models of axion inflation with a large axion decay constant in string theory. For models with a single axion, it was shown in \cite{Banks:2003sx} that it is impossible to obtain an axion decay constant larger than the Planck mass in many different examples. In fact, a general quantum gravity argument---the weak gravity conjecture (WGC) \cite{ArkaniHamed:2006dz}---suggests that this is not a coincidence but due to a general principle in string theory. According to the WGC, field excursions parametrically larger than the Planck mass may simply not be allowed in EFTs consistent with quantum gravity (see also \cite{Conlon:2012tz}). The conjecture was originally motivated using the example of an Abelian gauge field but argued to apply more generally to systems involving $p$-form fields. An extension of the conjecture was analyzed in detail for the case of multiple U$(1)$'s in \cite{Cheung:2014vva}. Furthermore, a number of recent works studied the implications of the WGC for multi-axion systems \cite{Rudelius:2014wla, delaFuente:2014aca, Rudelius:2015xta, Montero:2015ofa, Brown:2015iha, Bachlechner:2015qja, Hebecker:2015rya, Brown:2015lia}.

In \cite{Montero:2015ofa}, it was shown that contributions to the scalar potential from gravitational instantons forbid a super-Planckian field range, confirming that the WGC indeed bounds the field excursion in models with one or several axions.
In \cite{Brown:2015iha}, the problem was analyzed from a different point of view. There, the authors found a bound on the axion field range by relating the setup to the more tractable case of U(1)'s using T-dualities (see also \cite{Rudelius:2015xta} for similar arguments).
Both \cite{Montero:2015ofa} and \cite{Brown:2015iha} also discussed a possible loophole by which models with one or several axions might be able to evade a bound on the field excursion. In \cite{Bachlechner:2015qja}, the authors advocated the same loophole and argued that gravitational instantons do in general not spoil trans-Planckian field excursions in models with many axions. However, \cite{Brown:2015iha, Brown:2015lia} also gave counter-arguments suggesting that the loophole is unlikely to be realized in string theory.
On the other hand, string theory constructions possibly evading a bound on the field excursion were discussed in \cite{Bachlechner:2014gfa, Hebecker:2015rya}.
It is therefore fair to say that it is still under debate whether or not the WGC implies a strict bound on the axion field range in general models of axion inflation.

On the other hand, it is well-known that generic ``bottom-up'' models of axion inflation do allow for a parametric enhancement of the field excursion.\footnote{By ``bottom-up'', we mean both phenomenological and string theory inspired models which may be oblivious to possible constraints from the WGC.} One way to achieve this is to introduce monodromies for the axions by switching on couplings to branes or fluxes. Roughly, this changes the topology of the axion moduli space from a circle into a spiral, thus allowing for an enhancement of the naive field range \cite{Silverstein:2008sg, McAllister:2008hb, Marchesano:2014mla, Blumenhagen:2014gta, Hebecker:2014eua}.
In the absence of monodromies, an enhancement of the axion decay constant can be achieved by alignment \cite{Kim:2004rp} in models with at least $N=2$ axions, albeit at the cost of having to tune the anomaly coefficients entering the instanton potential to rather large values (see, however, \cite{Shiu:2015uva, Shiu:2015xda} for a recently proposed  alternative without this requirement). If one allows the number of axions to be large, $N \gg 1$, the possibilities for enhancement are richer. Even without any alignment, one then finds a Pythagorean enhancement $f_\text{eff} \sim \sqrt{N}$ via the $N$-flation mechanism \cite{Dimopoulos:2005ac}.\footnote{Possible embeddings of $N$-flation into type IIB string theory were discussed in \cite{Grimm:2007hs}.} In \cite{Bachlechner:2014gfa}, a statistical analysis of string theory inspired multi-axion systems revealed that some amount of alignment is actually generic at large $N$, leading to a stronger scaling $f_\text{eff} \sim N$ or even $f_\text{eff} \sim N^{3/2}$. Finally, it was found in \cite{Choi:2014rja} that a moderate tuning of the anomaly coefficients at large $N$ can even lead to an exponentially strong enhancement $f_\text{eff} \sim \sqrt{N!} \,n^N$, with $n$ an $\mathcal{O}(1)$ number.\footnote{See also \cite{Higaki:2014pja, Grimm:2014vva, Tye:2014tja, Kappl:2014lra, Ben-Dayan:2014zsa, Cicoli:2014sva, Bachlechner:2014hsa, Gao:2014uha, Peloso:2015dsa} for other recent works on axion inflation.}

In view of these results, there are two main questions we want to address in this paper. First, we are interested in how the enhancement of the effective axion decay constant scales with $N$ in a completely general bottom-up model of multi-axion inflation (statistically likely in the string landscape or not) with an arbitrary number $P$ of instantons contributing to the scalar potential and an arbitrary amount of alignment. To this end,
we derive an analytic expression for $f_\text{eff}$ in terms of a recurrence relation and determine its scaling with $N$ depending on the dihedral angles and distances of the facets in the $N$-polytope which bounds the fundamental domain of the axion moduli space. These parameters then in turn depend on the number of instantons and the alignment. We thus find a variety of different regimes in which $f_\text{eff}$ obeys different scaling laws with respect to $N$, including regimes of no enhancement, power-law enhancement and exponential enhancement. As special cases, we recover the $\sqrt{N}$ enhancement of $N$-flation, the $N$ and $N^{3/2}$ regimes found in \cite{Bachlechner:2014gfa} and the exponential enhancement found in \cite{Choi:2014rja}. Our result is useful in that it puts into perspective the different large-$N$ behaviors found in these works. Specifically, we will show that, at large $N$, the regimes of power-law and exponential enhancement are only separated by an infinitesimal change in the dihedral angles of order $\mathcal{O}(1/N)$. Moreover, our general algorithm for $f_\text{eff}$ opens up the possibility of a broad study of the phenomenology of multi-axion models with arbitrary scalar potential.

Second, we want to address the question how the enhancement mechanisms discussed above are constrained by quantum gravity. In particular, we are interested in the conditions which a general bottom-up model would have to satisfy in order to be consistent with a possible bound on the field excursion.
To what extent arguments such as those in \cite{Brown:2015iha} also hold for models with monodromies is not sufficiently understood so far. Nevertheless, an obstruction to an infinite field excursion is that
tunneling processes become probable when the potential energy of the axion becomes too large \cite{Kaloper:2011jz}. In order for these effects to be suppressed, the axion should not start to roll down the potential too far away from the minimum, which translates into an upper bound for the field excursion.\footnote{Another bottleneck is to actually construct an explicit string compactification with stabilized moduli which realizes an EFT with the desired properties \cite{Blumenhagen:2014nba, Hebecker:2014kva}.}
In this paper, we study the field excursion in models without monodromies. Our general recurrence relation allows us to translate this problem into a bound on the number of dominant instanton terms that contribute to the scalar potential.\footnote{We refer to instantons as ``dominant'' if they are relevant for the scalar potential and bound the field excursion along the direction in field space to which they couple. This is to be contrasted with subleading (multi-)instantons whose contribution to the scalar potential is negligible.} Our main result is that, in order for the enhancement to converge to a finite value at large $N$, the number of dominant instantons would have to grow faster than quadratically and possibly even exponentially with $N$. For the most part, the arguments leading to this conclusion are purely geometric and therefore model-independent.

There are two possible interpretations of this result. If the WGC (or other quantum gravity constraints) bound the axion field excursion to be sub-Planckian, this implies that
models of large-field inflation with multiple axions lie in the swampland because their assumptions on the form of the scalar potential become inconsistent for sufficiently large $N$. However, as we will detail in Section \ref{constraints}, it would then be difficult to evade the conclusion that string compactifications with many axions must admit an enormous number of dominant instantons, which is not expected in a perturbatively controlled regime.
Assuming that perturbative models with many axions exist, a second interpretation of our result may therefore be more reasonable: it should be taken as evidence that models with many axions must be able to violate a possible bound on the field excursion. Our result would thus lend further credence to the point of view that a parametric enhancement of the effective axion decay constant is compatible with quantum gravity. We stress that this argument is independent of and complementary to the loopholes discussed in \cite{Montero:2015ofa, Brown:2015iha, Bachlechner:2015qja}.

This paper is organized as follows. In Section \ref{wgc}, we review the WGC and its possible implications for the field excursion in models of inflation with axions.
In Section \ref{feff}, we derive the enhancement of the effective axion decay constant for a general model with $N$ axions in terms of a recurrence relation, which depends on the dihedral angles and distances of the facets in the $N$-polytope bounding the axion moduli space.
In Section \ref{constraints}, we relate these parameters to the number of dominant instanton terms in the scalar potential and argue that it must grow faster than quadratically and possibly even exponentially with $N$ in order for the enhancement to converge at large $N$. We conclude in Section \ref{conclusions} with a discussion of our results.

\section{The weak gravity conjecture}
\label{wgc}

The WGC postulates that gravity must be the weakest force in any consistent 4D low-energy EFT \cite{ArkaniHamed:2006dz}. More precisely, considering an EFT containing gravity and a U(1) gauge field with coupling constant $g$, there must exist at least one charged particle with mass
\begin{equation}
m \lesssim g M_\text{p}. \label{wgc-u1}
\end{equation}
The strong version of the WGC holds if the particle satisfying this bound is also the lightest charged particle in the theory, while otherwise \eqref{wgc-u1} is referred to as the weak or mild version of the WGC.

The above bound tells us that, whenever we make the gauge coupling $g$ too small, new particles will become light and destroy the validity of the EFT we started with. This can be seen by applying the above bound also to magnetic monopoles with charge $1/g$. Demanding that their mass should at least be of the order $\Lambda/g^2$ then yields a bound $\Lambda \lesssim g M_\text{p}$ for the cutoff of the EFT. The strength of the gauge interaction can therefore not be made arbitrarily small or, in other words, gravity is the weakest force. The conjecture is motivated by black hole arguments involving remnants and a related argument that string theory does not allow global symmetries. In the limit $g \to 0$, the gauge symmetry would become a global symmetry. This should not be allowed, which is reflected by the fact that the scale beyond which new physics corrects the 4D EFT then goes to zero, $\Lambda \to 0$.

A number of related conjectures for theories consistent with quantum gravity was formulated in \cite{Ooguri:2006in}. In particular, it was argued there that the moduli space of a consistent theory must have a finite volume and that travelling an infinite distance along a geodesic in field space would be accompanied by the appearance of an infinite tower of extra light states, thus leading to a break-down of the theory. A simple example illustrating this behavior is the volume of a string compactification.
If it is taken too large, KK modes become light, while, if it is made too small, string states become light and we would have to revert to the T-dual description instead.

\begin{figure}[t]
\centering
\includegraphics[trim = 0mm 90mm 0mm 30mm, width=1\textwidth]{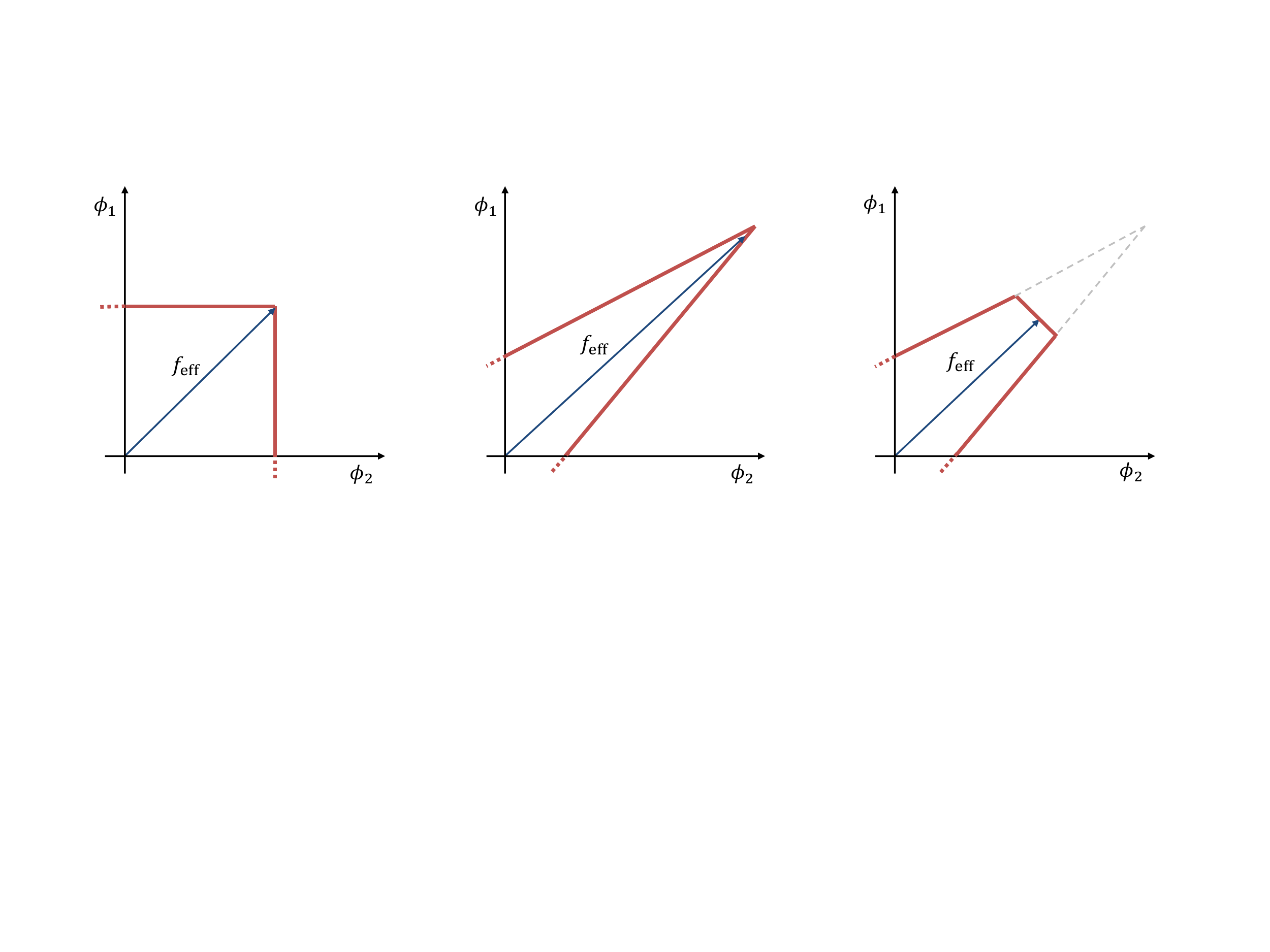}
\caption{The fundamental domain of the moduli space of two axions for $N$-flation, ``naive'' alignment and alignment including quantum gravity constraints. $N$-flation yields an effective axion decay constant $f_\text{eff}$ enhanced by a factor of $\sqrt{2}$, which is compatible with the WGC as long as $f_\text{eff}$ is not super-Planckian. From the bottom-up perspective, alignment can lead to a very large enhancement. However, unless axions can exploit a loophole, the WGC predicts that new terms then become relevant in the scalar potential and shorten the field range.\label{fig-wgc}}
\end{figure}

It was also argued in \cite{ArkaniHamed:2006dz} that the WGC can be generalized to $p$-form fields and $(p-1)$-dimensional objects charged under them. Since an axion has $p=0$, the natural objects to consider in this case are instantons, where the role of the gauge coupling is played by the inverse axion decay constant $1/f$. Instantons correct the scalar potential of an axion by terms of the form
\begin{equation}
V(\phi) \sim \e^{-S_\text{E}} \left[1-\cos\left(\frac{\phi}{f}\right)\right], \label{scalar}
\end{equation}
where the suppression of the correction is controlled by the size of the Euclidean action $S_\text{E}$. Hence, $S_\text{E}$ is the analogue to the mass $m$ in \eqref{wgc-u1}, which determines whether the charged particle is heavy enough to be integrated out in the low-energy EFT.
A natural guess for a generalization of the bound \eqref{wgc-u1} to axions is therefore
\begin{equation}
S_\text{E} \lesssim \frac{M_\text{p}}{f}, \label{wgc-axions}
\end{equation}
i.e., for any axion with axion decay constant $f$, at least one instanton should exist satisfying the above bound. This implies that, whenever the axion decay constant is taken larger than the Planck mass, we expect large corrections to the scalar potential \eqref{scalar} from instantons satisfying \eqref{wgc-axions} and their higher harmonics, spoiling a super-Planckian field range (see Fig. \ref{fig-wgc}). Indeed, such large instanton corrections were the main obstacle in \cite{Banks:2003sx} to constructing large-field models with a single axion. That \eqref{wgc-axions} is the correct way of generalizing the WGC to axions was recently also shown in \cite{Montero:2015ofa} via an explicit computation of gravitational instantons and in \cite{Brown:2015iha} by relating axions to U(1) gauge fields using T-dualities.

A possible loophole to the bound \eqref{wgc-axions} was proposed in \cite{Brown:2015iha}, where it was argued that the instantons satisfying the bound might not be those with the smallest action (which is admissible if only the weak version of the WGC holds). These instantons could therefore be suppressed, while other instantons, which do not satisfy the WGC bound, would give the dominant contributions, thus allowing for a super-Planckian field excursion. The same proposal was also advocated in \cite{Bachlechner:2015qja}. In a similar spirit, it was argued in \cite{Montero:2015ofa} that the presence of a discrete gauge symmetry can forbid the existence of some instantons in models with $N$ axions such that the true bound on $S_\text{E}$ can be larger than \eqref{wgc-axions} along some directions in field space. In \cite{Bachlechner:2015qja}, it was furthermore argued that such a larger bound is even generic at large $N$. The true bound would then be
\begin{equation}
S_\text{E} \lesssim N^\lambda\frac{M_\text{p}}{f}
\end{equation}
with a model-dependent exponent $\lambda>0$. If this is correct, it should be possible to engineer models with an effective axion decay constant $f_\text{eff}\sim N^\lambda M_\text{p}$ along some diagonal in the axion field space and, hence, obtain super-Planckian field excursions.
However, it was argued in \cite{Brown:2015iha, Brown:2015lia} that it is unlikely that this loophole can be realized in a consistent string theory model.

Another proposed loophole \cite{delaFuente:2014aca} is that, in certain models, instanton corrections may come with extra suppression factors such that they can be suppressed in the scalar potential even if their Euclidean action is small. In such models, one might be able to satisfy the strong WGC and still obtain super-Planckian field ranges. However, an explicit realization of this loophole has not been obtained so far.

To summarize, it has not been fully understood so far whether or not super-Planckian field excursions are allowed by the WGC.
Given that future observations might reveal a large tensor-to-scalar ratio, it is of course crucial to settle this issue.
It should therefore be worthwhile to also analyze the problem from a different point of view, as we will do in the following sections.

\section{The effective axion decay constant}
\label{feff}

In this section, we compute the enhancement of the effective axion decay constant for a completely general model with $N$ axions and $P$ instanton terms in the potential. The Lagrangian of such a model is of the general form
\begin{equation}
\mathcal{L}(\phi_i) = \frac{1}{2} \sum^N_{i=1} (\partial \phi_i)^2 + \sum^P_{j=1} \Lambda^4_j \left[1-\cos\left( \sum^N_{i=1} c_{ij}\phi_i\right)\right], \label{a}
\end{equation}
where $\Lambda_j, c_{ij}$ are numerical constants and we have chosen a field basis in which the axions are canonically normalized.
The periodicity of the instanton potential implies that the fundamental domain of the $N$-dimensional axion moduli space is an $N$-polytope with $2P$ facets \cite{Bachlechner:2014gfa}. For $N=2$, this would be a polygon with $2P$ edges, for $N=3$, it would be a polyhedron with $2P$ faces, etc. Our goal in this section is to study the enhancement in the direction of one vertex of such a general polytope. In Section \ref{feff-1}, we will first derive this for a simplified model in which the different facets intersecting at the vertex have equal dihedral angles and distances from the origin. We will see that this simple toy model already gives us a good qualitative understanding of the scaling behavior of the enhancement at large $N$. In Section \ref{feff-2}, we will then generalize our results to a completely general polytope.

Our approach is somewhat different from the one employed in \cite{Bachlechner:2014gfa}, which also studied enhancement in models of the form \eqref{a}. First, we will not use landscape statistics in this section but compute the field excursion for an arbitrary polytope, statistically likely in the string landscape or not.
Second, we will determine the enhancement in terms of dihedral angles.
We will see that the scaling behavior of the enhancement at large $N$ is largely determined by the expectation value of the dihedral angles at the corresponding vertex and tends to be further increased when variations are introduced.
Furthermore, our approach will allow us to relate the axion field range to the number of instantons $P$ in Section \ref{constraints}.

\subsection{A simple model}
\label{feff-1}

Let us now analyze the enhancement in the direction of a particular vertex for which the $N$ $(N-1)$-facets intersecting at the vertex all have the same dihedral angles and the same distances from the origin.\footnote{In principle, more than $N$ facets can intersect at a vertex of an $N$-polytope. This is ungeneric and does not lead to further enhancement compared to the generic case of $N$ facets.}
We denote the normal vectors pointing from the origin to the facets by $\vec f^{(i)}_1$ with $i=1,\ldots,N$. By assumption, all $\vec f^{(i)}_1$ have the same length $f_1 < M_\text{p}$.
Let us also assume that the angle between each pair of the normal vectors is the same and denote this angle by $\alpha_2$. By slight abuse of terminology, we will refer to $\alpha_2$ as the dihedral angle in this paper.\footnote{Actually, the dihedral angle is $\pi-\alpha_2$.}
Let us furthermore denote vectors normal to $(N-n)$-facets by $\vec f_n$ and their lengths by $f_n$. Consequently, the vector pointing towards the vertex is denoted by $\vec f_N$ and its length by $f_N$.\footnote{It is conventional in the literature to define the axion decay constant $f$ as $\frac{1}{2\pi}$ times the length between two maxima of the cosine potential, which would correspond to twice the distance between a facet and the origin. Deviating from this convention saves us factors of $\pi$ in numerous expressions. For the main quantity of interest, the enhancement $f_N/f_1$, our convention has no effect. Also note that the definition of an axion decay constant is somewhat ambiguous in models with more than one axion \cite{Bachlechner:2014gfa, Bachlechner:2015qja}. In this paper, the term ``axion decay constant'' always refers to distances of facets from the origin, while ``effective axion decay constant'' or ``enhancement of the axion decay constant'' refers to distances of vertices from the origin.}
As an example, consider the case $P=N$, where, in the absence of alignment, the fundamental domain is an $N$-cube. The normal vectors are then orthogonal to each other such that $\alpha_2=\frac{\pi}{2}$, and the enhancement of the effective axion decay constant grows like $f_N/f_1 = \sqrt{N}$. Our goal is now to work out the enhancement for general angles $\alpha_2 \neq \frac{\pi}{2}$.

\begin{figure}[t]
\centering
\includegraphics[trim = 0mm 60mm 0mm 10mm, width=1\textwidth]{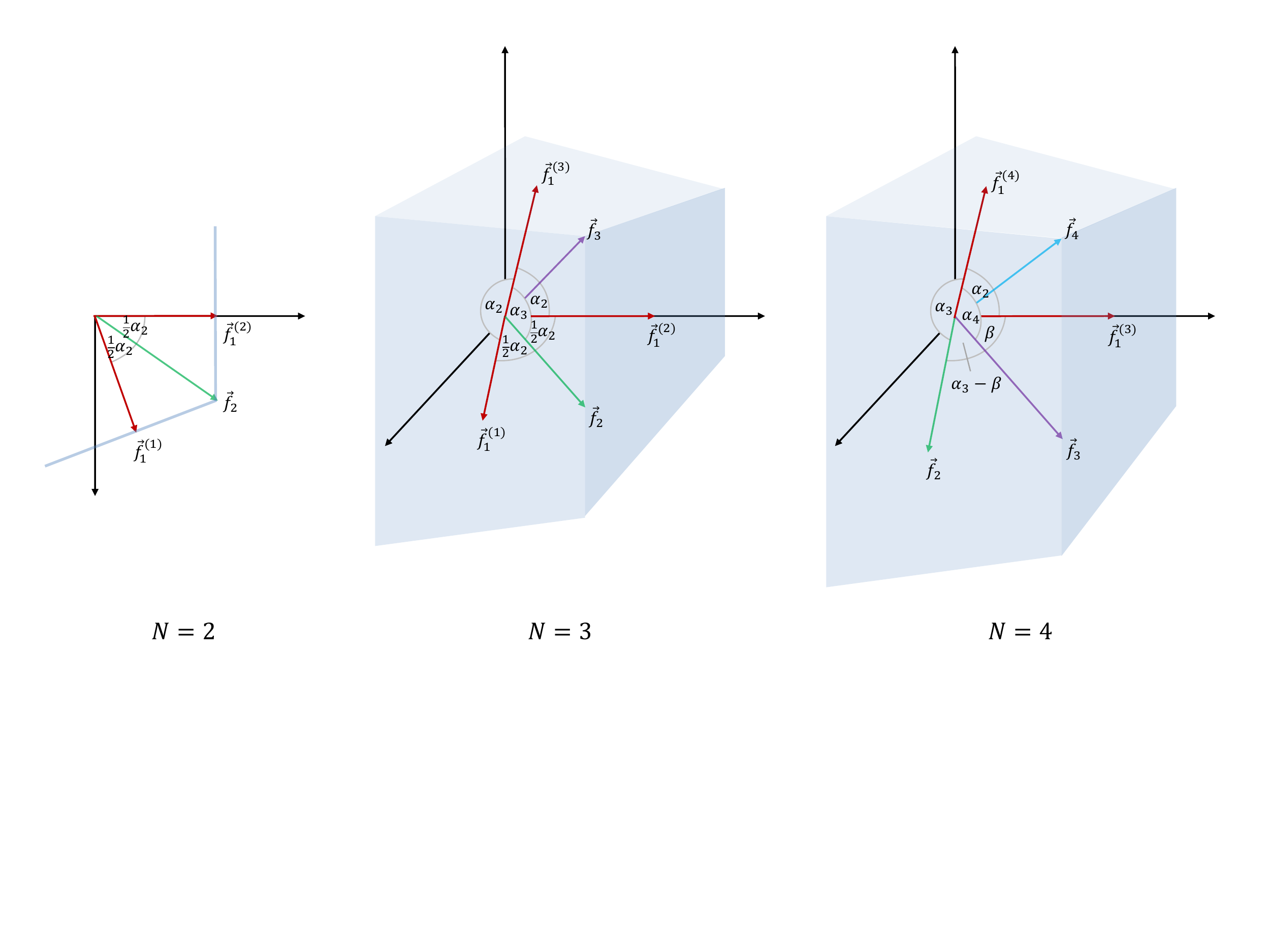}
\caption{Normal vectors and the angles between them in a simple model for $N=2$, $N=3$ and $N=4$.\label{fig-polytope-one-angle}}
\end{figure}

We start with the case $N=2$. Then, at each vertex, two facets intersect such that we have two normal vectors $\vec f^{(1)}_1$ and $\vec f^{(2)}_1$. We can read off from Fig. \ref{fig-polytope-one-angle} that the enhancement is then given by
\begin{equation}
f_2 = \frac{f_1}{\cos \left( \frac{1}{2}\alpha_2\right)}.
\end{equation}

For $N=3$, we have three normal vectors, all with relative angles $\alpha_2$. We can now compute $f_3$ recursively from $f_2$ and $f_1$ as follows. For convenience, we choose our coordinate system such that $\vec f^{(1)}_1$, $\vec f^{(2)}_1$ and $\vec f_2$ lie in the $(\vec x,\vec y)$-plane (see Fig. \ref{fig-polytope-one-angle}). We can then compute scalar products between the different vectors in order to determine the components of $\vec f^{(3)}_1$ in terms of $\alpha_2$. This in turn allows us to compute the angle between $\vec f^{(3)}_1$ and $\vec f_2$, which we denote by $\alpha_3$. We find
\begin{equation}
\alpha_3 = \arccos \left[ \frac{2\cos^2 \left( \frac{1}{2}\alpha_2\right) -1}{\cos \left( \frac{1}{2}\alpha_2\right)} \right].
\end{equation}
Furthermore, one can verify that $\vec f^{(3)}_1$ lies in the plane spanned by $\vec z$ and $\vec f_2$. This follows from symmetry (see again Fig. \ref{fig-polytope-one-angle}): flipping the polyhedron along the $(\vec z,\vec f_2)$-plane has the effect that $\vec f^{(1)}_1$ and $\vec f^{(2)}_1$ are exchanged. However, since the angle between $\vec f^{(3)}_1$ and $\vec f^{(1)}_1$ and the angle between $\vec f^{(3)}_1$ and $\vec f^{(2)}_1$ are equal, $\vec f^{(3)}_1$ must be invariant under the flip. Hence, it lies in the $(\vec z,\vec f_2)$-plane and its length along this plane is given by its absolute value $f_1$. This implies $f_3 = f_1/\cos \beta$, which together with $f_3 = f_2/\cos (\alpha_3-\beta)$ yields $f_3 = f_1/\cos\big(\alpha_3-\arccos\big(\frac{f_2}{f_3}\big)\big)$. The last equation can now be solved for $f_3$, with the result
\begin{equation}
f_3 = \frac{\sqrt{f_1^2+f_2^2 - 2 \cos (\alpha_3)f_1 f_2}}{\sin (\alpha_3)}.
\end{equation}

It is straightforward to continue this iterative process for higher $n=4,\ldots,N$. The general recurrence relation is then
\begin{equation}
f_n = \frac{\sqrt{f_1^2+f_{n-1}^2 - 2 \cos (\alpha_n)f_1 f_{n-1}}}{\sin (\alpha_n)}, \label{rec2}
\end{equation}
where the angle is given by
\begin{equation}
\alpha_n = \arccos \left[ \sqrt{1- \frac{f_1^2}{f_{n-1}^2}} \frac{1-\cos(\alpha_2)}{\tan (\alpha_{n-1})}+ \frac{f_1}{f_{n-1}} \cos (\alpha_2) \right].
\end{equation}

Let us now analyze in detail the different regimes of enhancement as we vary the dihedral angle $\alpha_2$.
We first check what happens when we deviate infinitesimally from the $N$-cube case, i.e., we consider $\alpha_2 = \frac{\pi}{2} + a \epsilon$. Computing $\alpha_n$ for the first few orders, we then find
\begin{equation}
\alpha_n = \frac{\pi}{2} + \sqrt{n-1} a \epsilon.
\end{equation}
In the limit $\epsilon\to 0$, we must recover the Pythagorean enhancement. We can therefore make the ansatz $f_n = \sqrt{n}f_1 + c_n f_1 a \epsilon$ for some coefficient $c_n$. Using these expressions in \eqref{rec2} and expanding in powers of $\epsilon$, we find
\begin{equation}
c_n = \sqrt{n} - \frac{1}{\sqrt{n}} + \sqrt{\frac{{n-1}}{n}} c_{n-1}.
\end{equation}
For large $n$, this becomes $c_n \approx \sqrt{n} + c_{n-1}$ and, hence, $c_N \sim N^{3/2}$. This implies that our expansion of $f_n$ already breaks down for $\epsilon \gtrsim \mathcal{O}(1/N)$. Hence, we expect to leave the regime of Pythagorean enhancement for all $\alpha_2$ which deviate from $\frac{\pi}{2}$ by at least this magnitude. For $\alpha_2 < \frac{\pi}{2}$, the enhancement will be slower than the $\sqrt{N}$ law, while it will be faster for $\alpha_2 > \frac{\pi}{2}$.

\begin{figure}[p]
\centering
\includegraphics[width=0.9\textwidth]{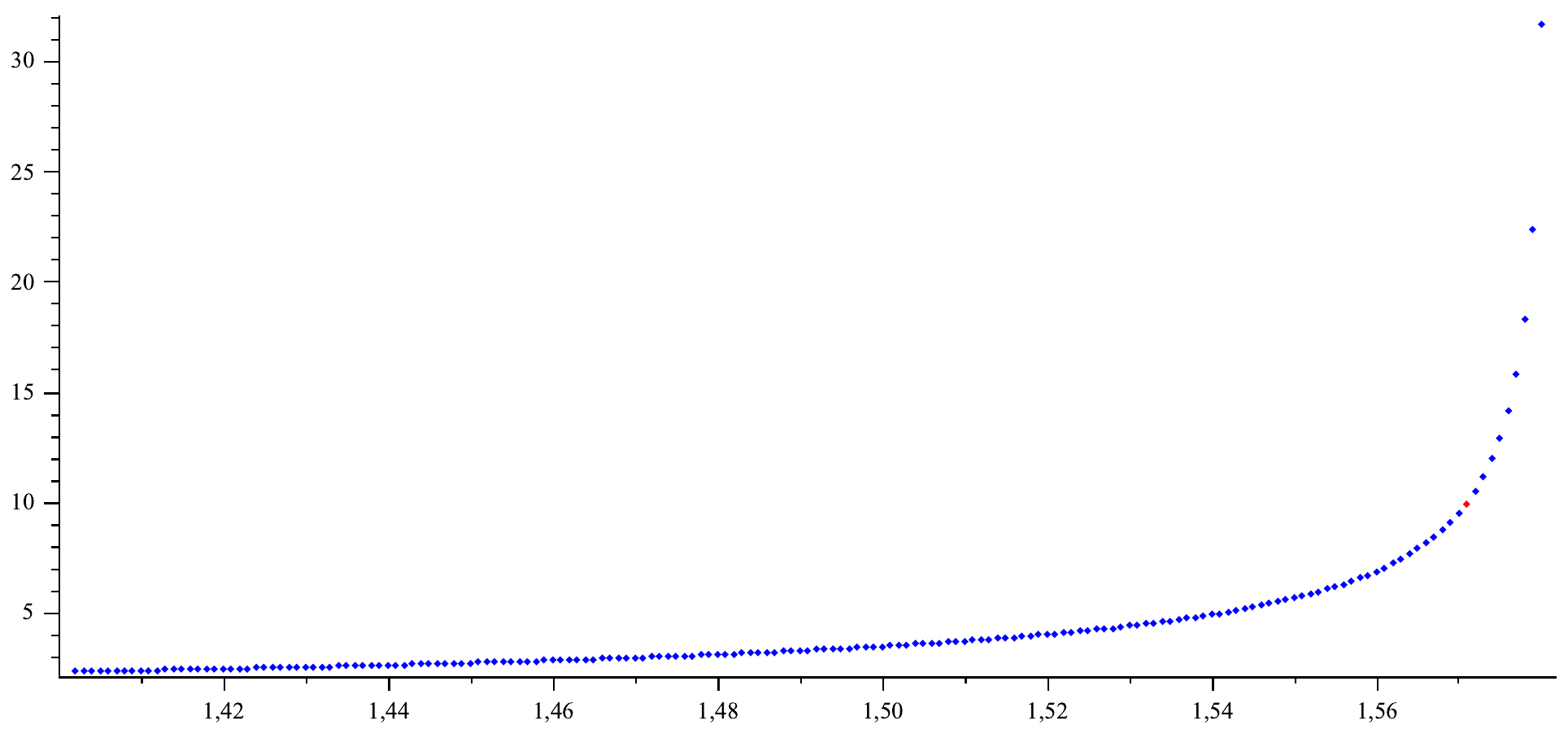}
\put(-430,100){$\frac{f_{10^2}}{f_1}$}
\put(-200,-10){$\alpha_2$}\\[1em]
\includegraphics[width=0.9\textwidth]{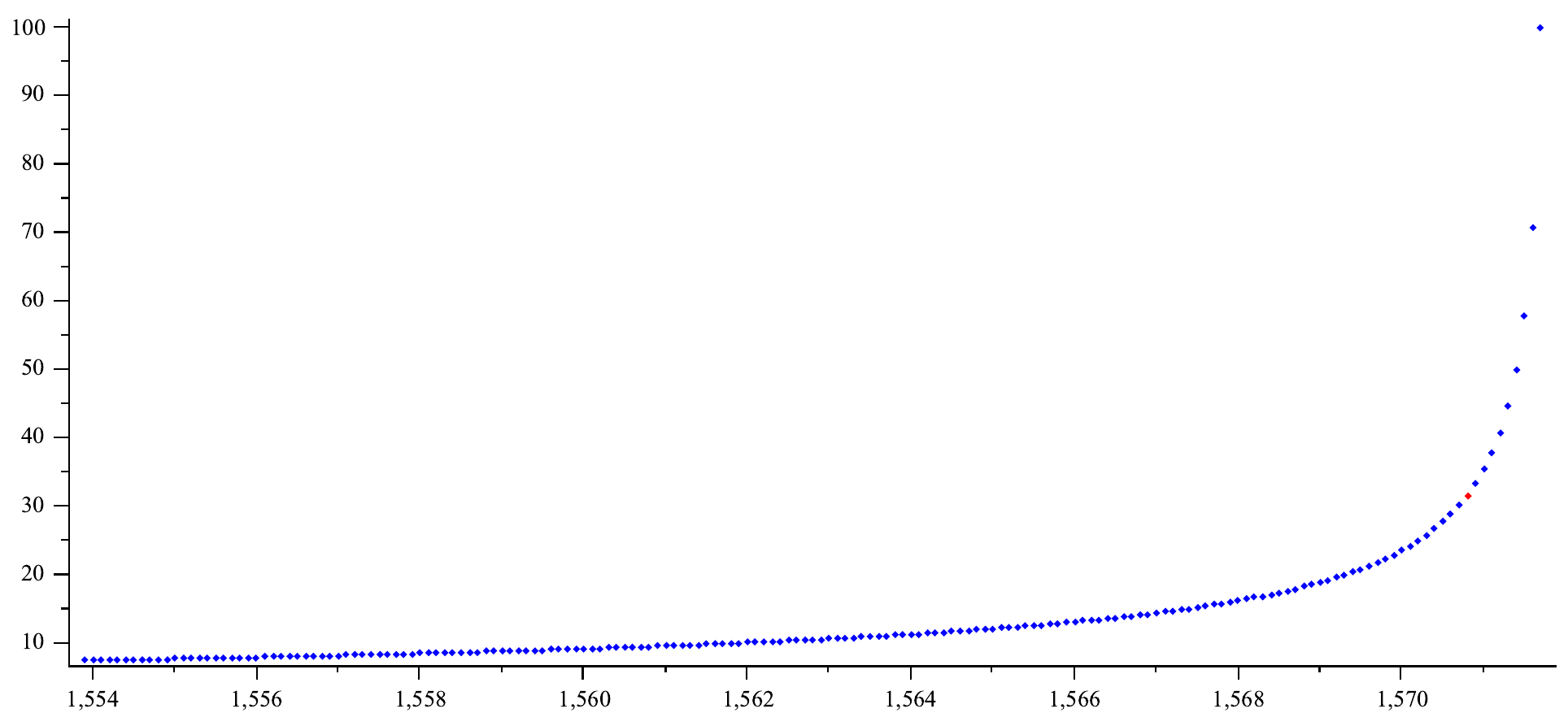}
\put(-430,95){$\frac{f_{10^3}}{f_1}$}
\put(-200,-10){$\alpha_2$}\\[1em]
\caption{The enhancement of the effective axion decay constant for $N=10^2$ and $N=10^3$ and different choices for the dihedral angle $\alpha_2$. The enhancement is slow for angles $\alpha_2 < \frac{\pi}{2}$ and diverges for angles $\alpha_2 > \frac{\pi}{2}$, where the different regimes are separated by $\Delta \alpha_2 \sim \mathcal{O}(1/N)$. The red dots denote the $N$-flation case $\alpha_2=\frac{\pi}{2}$. \label{fig-fixedN}}
\end{figure}

\begin{figure}[p]
\centering
\includegraphics[width=0.9\textwidth]{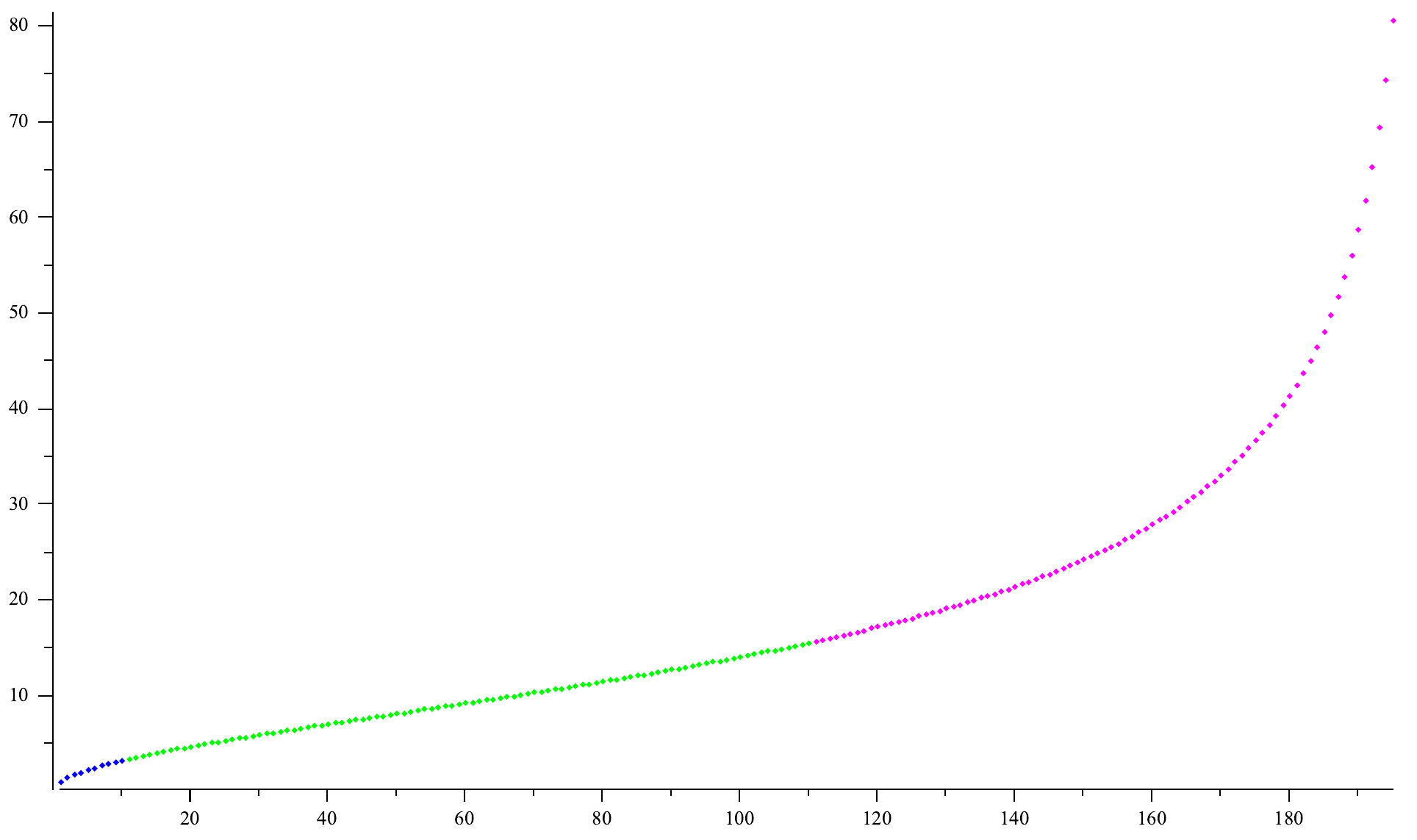}
\put(-425,125){$\frac{f_N}{f_1}$}
\put(-200,-10){$\scriptstyle{N}$}\\[1em]
\includegraphics[width=0.27\textwidth]{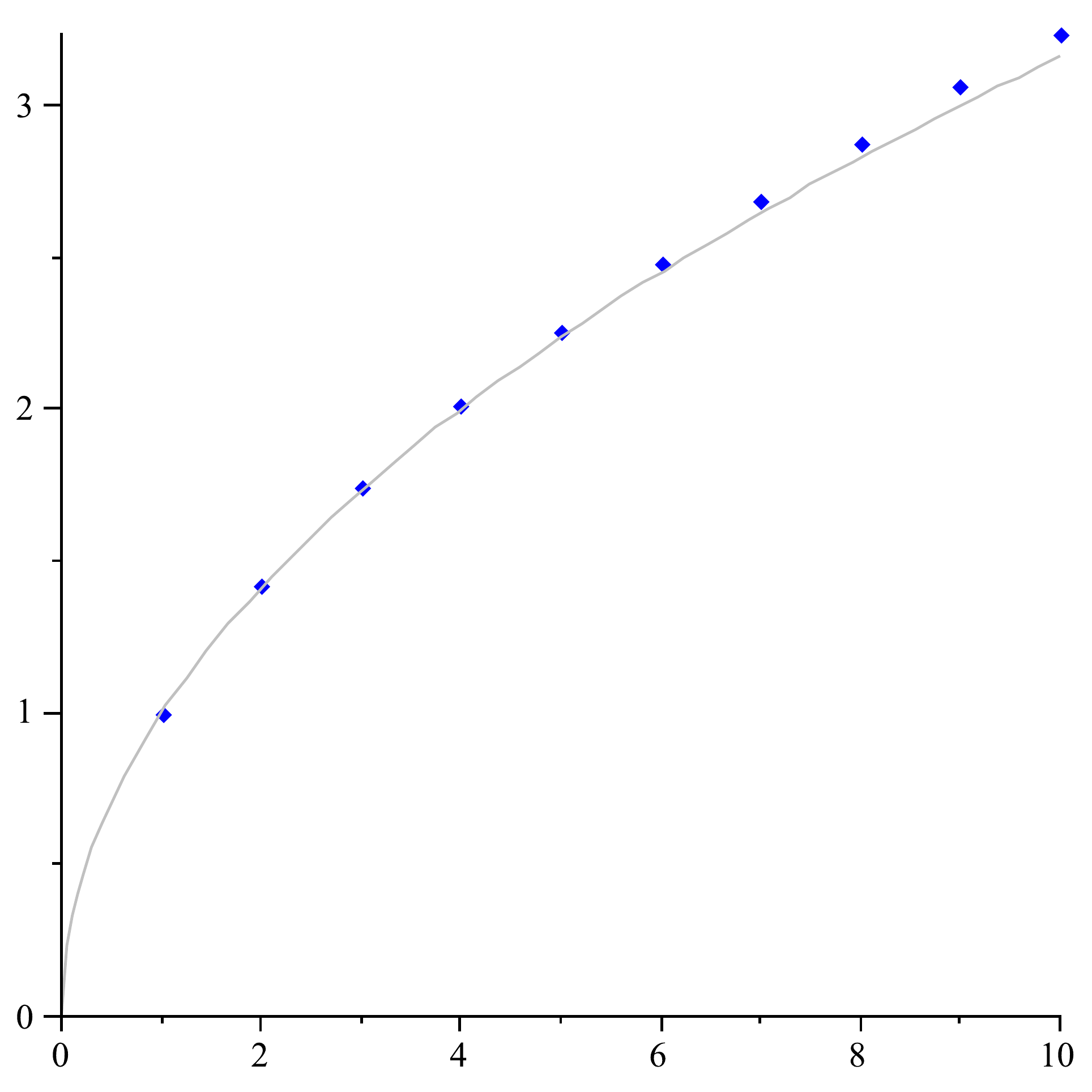}\qquad\includegraphics[width=0.27\textwidth]{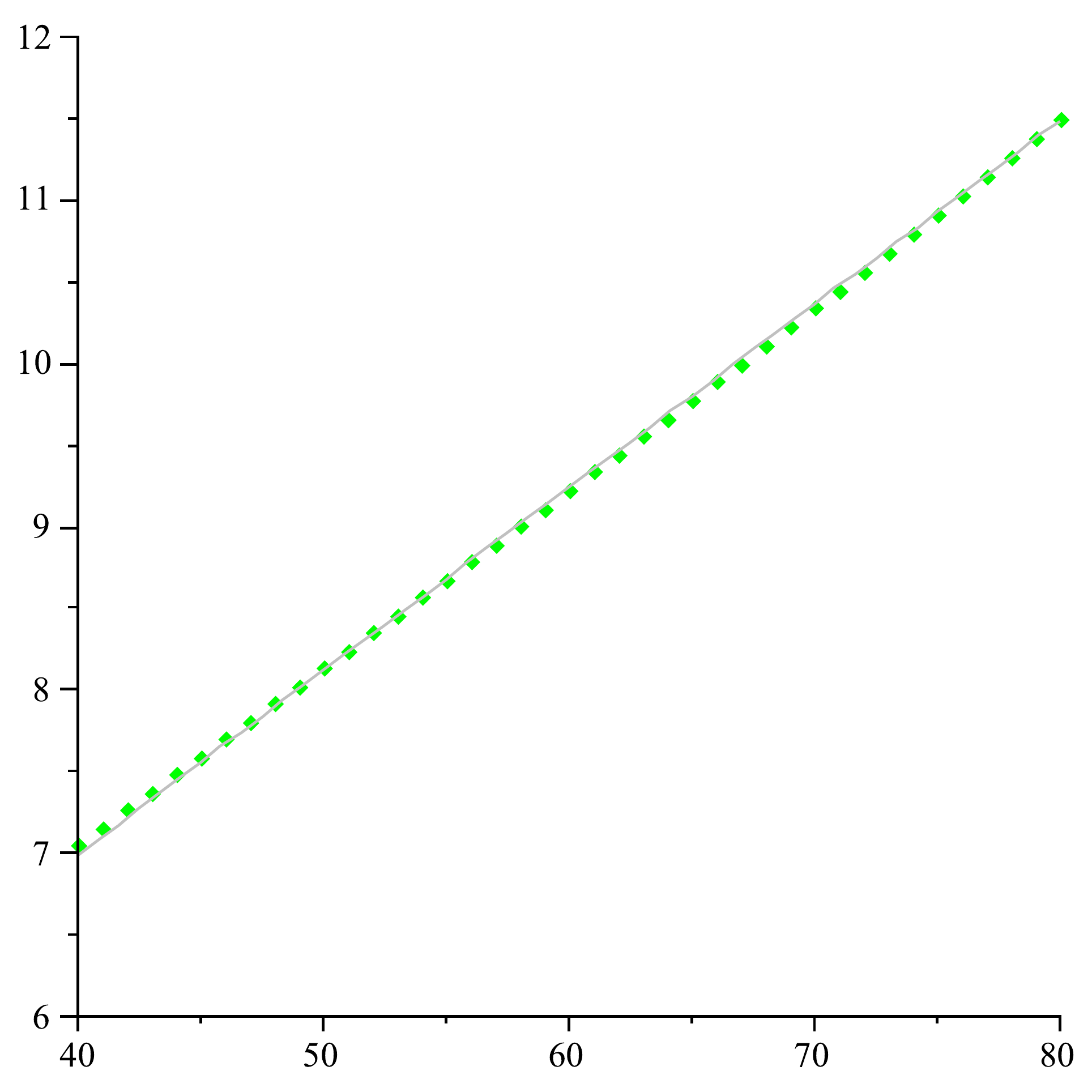}\qquad\includegraphics[width=0.27\textwidth]{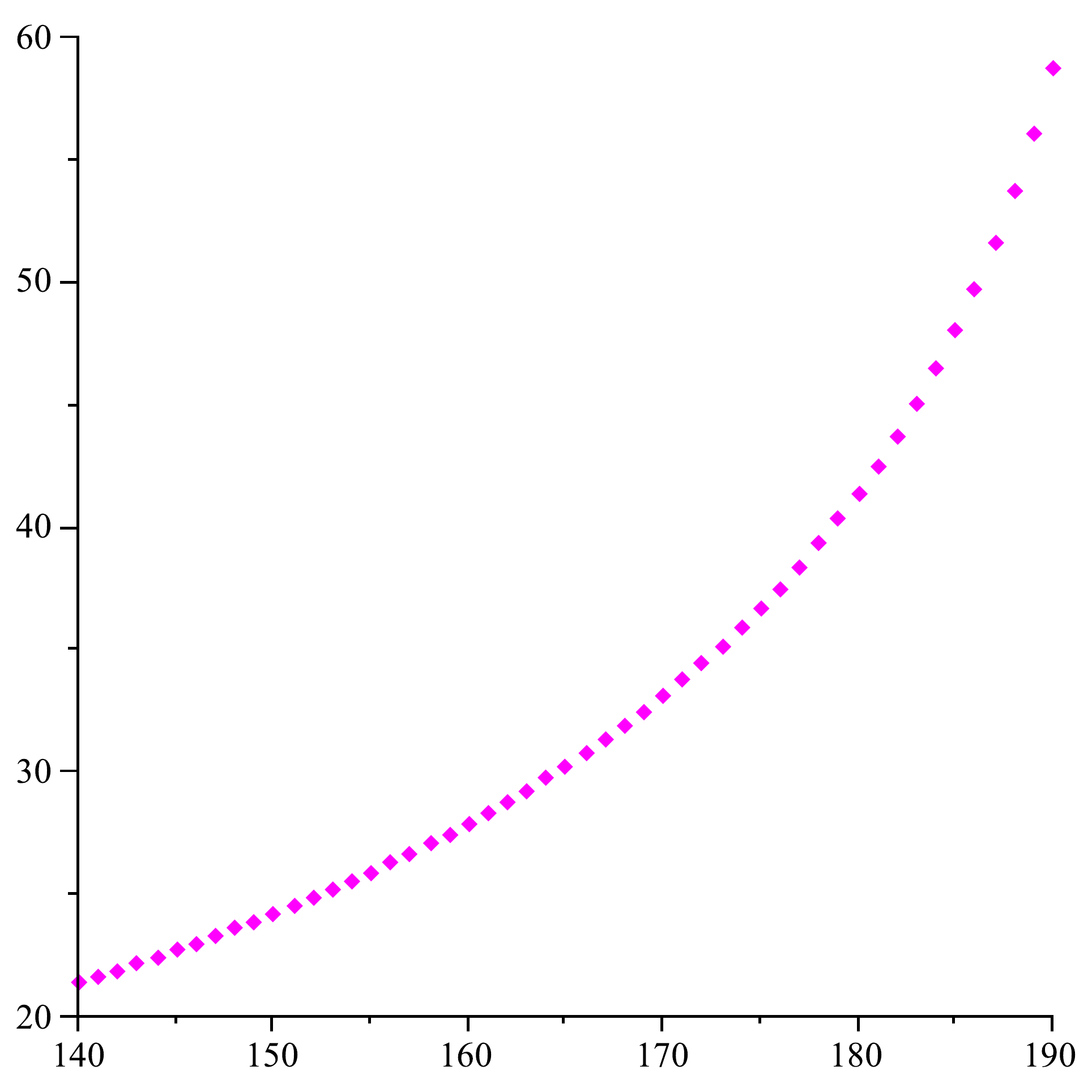}
\caption{The enhancement of the effective axion decay constant for fixed angle $\alpha_2= \frac{\pi}{2}+ \frac{1}{200}$ and different $N$. For small $N \ll 200$, we recover the Pythagorean $\sqrt{N}$ law (blue) since the deviation of $\alpha_2$ from $\frac{\pi}{2}$ is smaller than $1/N$. For larger $N$, the enhancement starts to grow linearly (green). Close to the limit $N = 200$ where $\alpha_2-\frac{\pi}{2} = 1/N$, the enhancement grows polynomially and later (super-)exponentially (violet). Note that there is nothing special to the value $N=200$: had we chosen a smaller (larger) angle, the divergence would have appeared at larger (smaller) $N$.\label{fig-fixedalpha}}

\end{figure}

\begin{figure}[t]
\centering
\includegraphics[width=0.9\textwidth]{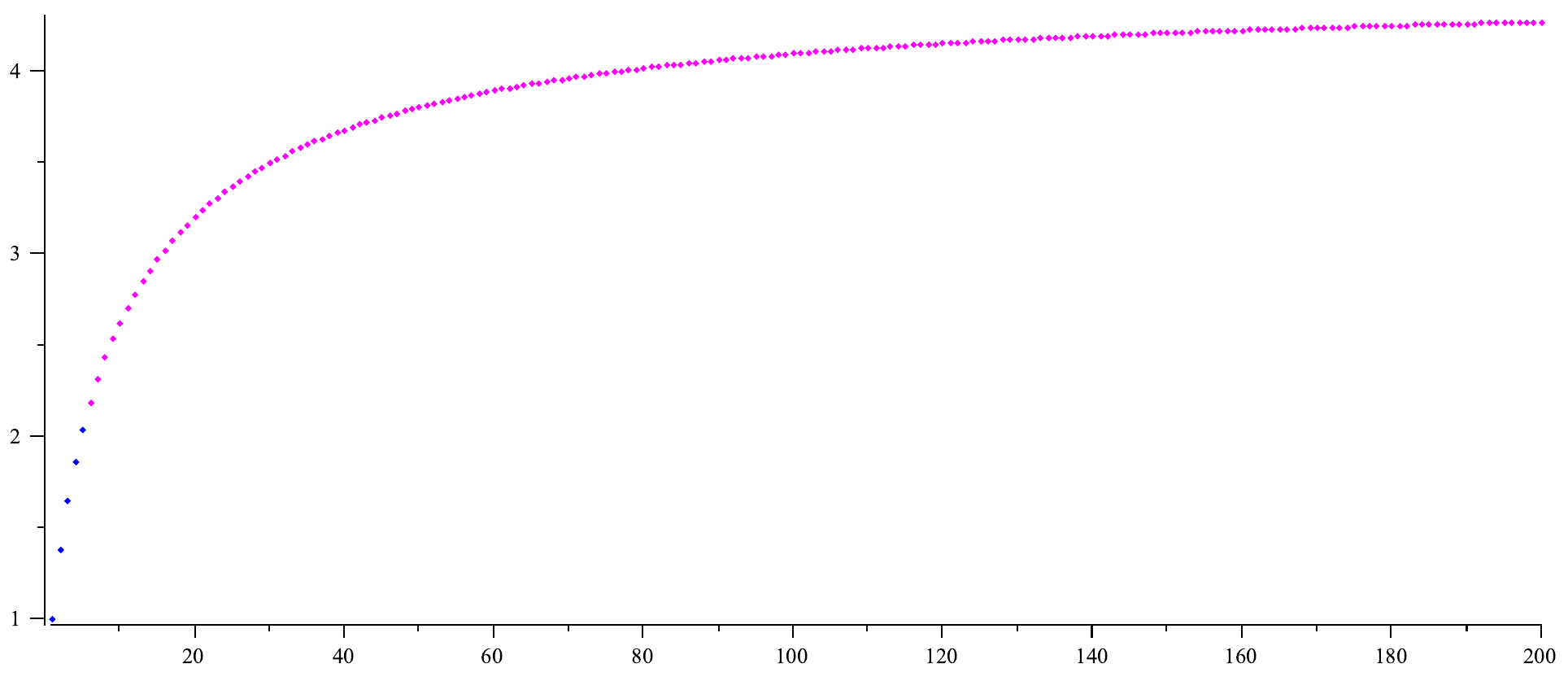}
\put(-425,90){$\frac{f_N}{f_1}$}
\put(-200,-10){$\scriptstyle{N}$}\\[1em]
\caption{The enhancement of the effective axion decay constant for fixed angle $\alpha_2= \frac{\pi}{2}- \frac{1}{20}$ and different $N$. For small $N \ll 20$, we recover the Pythagorean $\sqrt{N}$ law (blue), while for larger $N$ the enhancement grows more slowly and finally dies out completely (violet).\label{fig-fixedalpha-5}}

\end{figure}

Moreover, there is in fact an upper bound for $\alpha_2$ at which the enhancement diverges. That such a bound exists is easy to see in two dimensions. There, a vertex is determined by two normal vectors, which can at most differ by an angle of 180$^\circ$, i.e., $\alpha_2 \le \pi$. When the inequality is almost saturated, the two facets are almost parallel, and the alignment becomes infinitely large. In three dimensions, three normal vectors that each differ by the same angle can at most differ by 120$^\circ$. By computing scalar products of $N$ unit vectors in $N$ dimensions and demanding that they are all equal to $\cos (\alpha_2)$, we find the general rule,
\begin{equation}
\alpha_2 \le \pi - \arccos \left(\frac{1}{N-1}\right).
\end{equation}
For large $N$, this becomes
\begin{equation}
\alpha_2 \le \frac{\pi}{2} + \frac{1}{N}.
\end{equation}
Near this angle, the enhancement diverges, while we approach the Pythagorean regime for $\alpha_2 \to \frac{\pi}{2}$.

The different regimes of enhancement can be observed in Fig. \ref{fig-fixedN}, where we plotted the effective axion decay constant for different angles $\alpha_2$ at fixed $N$. 
The plot shows that the enhancement slows down compared to the $\sqrt{N}$ law when $\alpha_2$ is smaller than $\frac{\pi}{2}$ by at least an $\mathcal{O}(1/N)$ number and dies out completely as soon as $\alpha_2$ is sufficiently far away from $\frac{\pi}{2}$. If, on the other hand, $\alpha_2$ is larger than $\frac{\pi}{2}$ by $\mathcal{O}(1/N)$, the enhancement becomes huge and diverges.
Note that this also implies that, holding $\alpha_2\neq \frac{\pi}{2}$ fixed and increasing $N$, one is bound to leave the Pythagorean regime in either of the two directions for sufficiently large $N$. This is plotted in Figs. \ref{fig-fixedalpha} and \ref{fig-fixedalpha-5}, where one can nicely observe the different scaling laws with respect to $N$ as one scans through the different enhancement regimes.

To summarize, we have seen that, for large $N$, infinitesimal deviations $|\alpha_2-\frac{\pi}{2}| \sim \mathcal{O}(1/N)$ are sufficient to move away from the regime of Pythagorean enhancement, either to a regime of smaller enhancement or to one of an extremely large enhancement. Which one of the two possibilities applies depends on the angle $\alpha_2$ on the vertex in question, which is determined by two competing effects:
\begin{itemize}
 \item Increasing the number of dominant instantons $P$: this decreases the angles on some or all vertices of the polytope.
 \item Alignment: this increases the angles on some vertices and decreases it on others.
\end{itemize}
Hence, whether there is large enhancement on a particular vertex depends on the interplay between these two effects. If we add a lot of terms to the scalar potential such that $P \gg N$ but only admit little or no alignment, we will have $\alpha_2 < \frac{\pi}{2}$ and little or no enhancement. If, on the other hand, we take $P$ to be equal to or slightly larger than $N$ and at the same time allow strong alignment, we will get $\alpha_2 > \frac{\pi}{2}$ and the enhancement will be huge.\footnote{This seems to be at odds with the results of \cite{Bachlechner:2014gfa}, where a large enhancement was found for large $N$ \emph{independent} of the number of instantons $P$. This discrepancy is presumably due to the fact that the claims of \cite{Bachlechner:2014gfa} are of a statistical nature. Hence, their result should be interpreted such that it is statistically likely in the string landscape that the effect of alignment dominates over the effect of an increased number of constraints. This does not contradict our above statement that a regime of no enhancement may be reached for sufficiently large $P$ and sufficiently small alignment.} Of course, without making quantitative the relation between the dihedral angles and physical quantities such as the number of instanton terms or the anomaly coefficients, the above results are just geometry and do not immediately improve our understanding of actual string theory models of inflation. We will study the relation between the enhancement and the instantons in more detail in Section \ref{constraints}.

\subsection{General polytopes}
\label{feff-2}

\begin{figure}[p]
\centering
\includegraphics[width=0.9\textwidth]{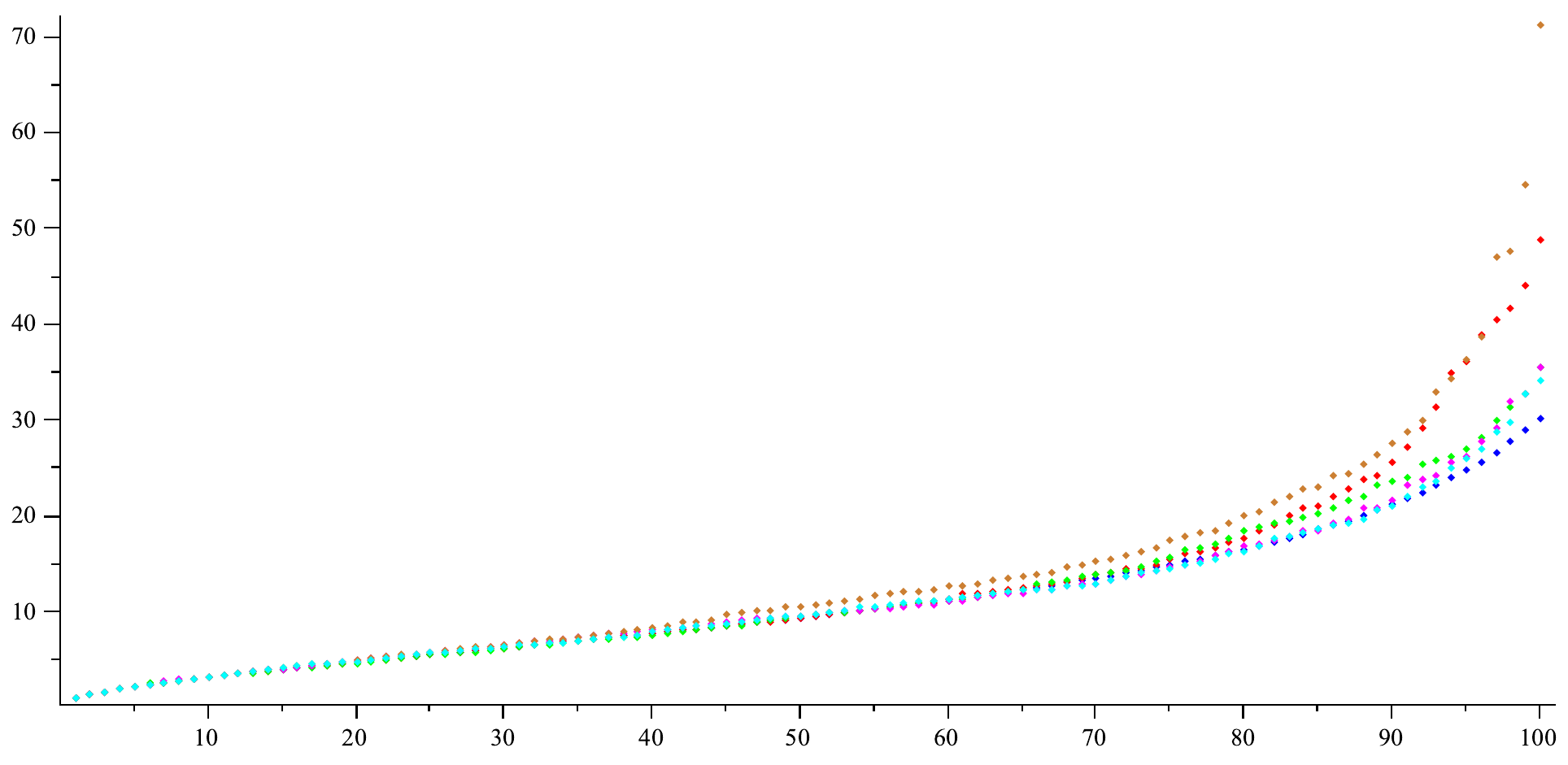}
\put(-425,110){$\frac{f_N}{f_1}$}
\put(-200,-10){$\scriptstyle{N}$}\\[1em]
\includegraphics[width=0.9\textwidth]{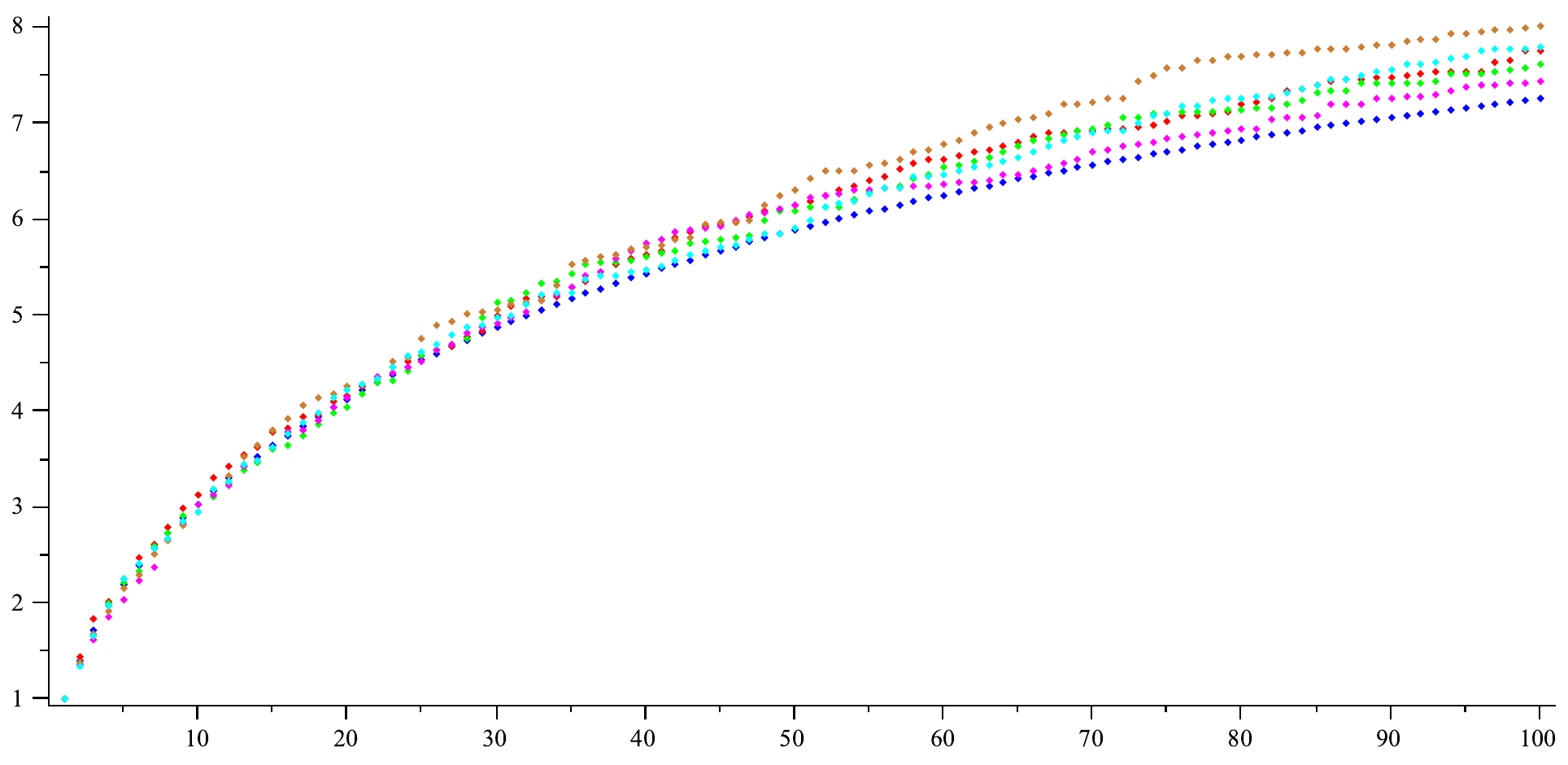}
\put(-425,110){$\frac{f_N}{f_1}$}
\put(-200,-10){$\scriptstyle{N}$}\\[1em]
\caption{The enhancement $f_N/f_1$ for $\langle\alpha_2\rangle = 0.58 > \frac{\pi}{2}$ (upper plot) and $\langle\alpha_2\rangle = 0.56 < \frac{\pi}{2}$ (lower plot).
Each plot for $\sigma(\alpha_2) = 0$ (blue dots) and five realizations of a uniform distribution with $\sigma(\alpha_2) = 0.029$ (upper plot) or $\sigma(\alpha_2) = 0.058$ (lower plot).\label{fig-random-poly-1}}
\end{figure}

\begin{figure}[p]
\centering
\includegraphics[width=0.9\textwidth]{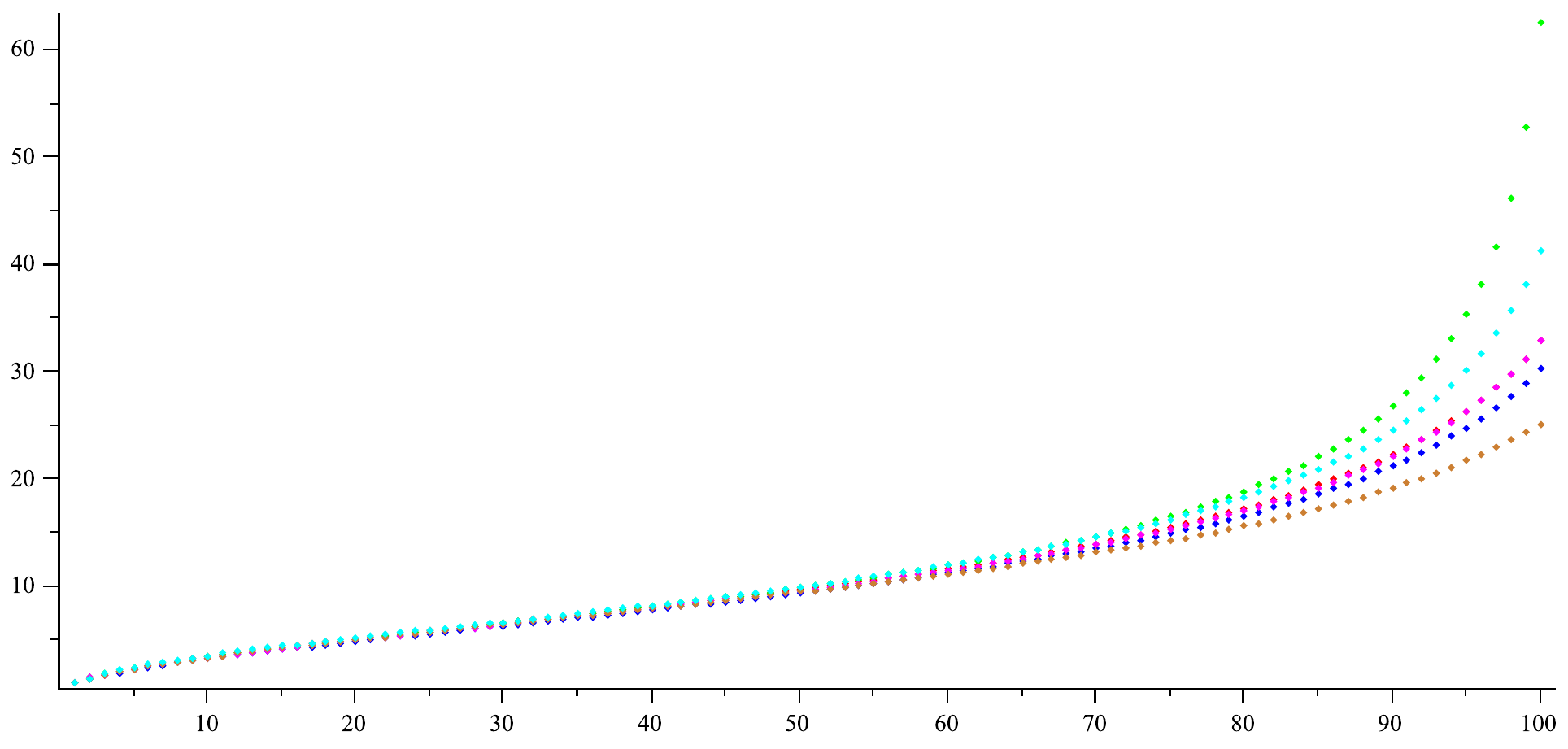}
\put(-425,110){$\frac{f_N}{\langle f_1\rangle}$}
\put(-200,-10){$\scriptstyle{N}$}\\[1em]
\includegraphics[width=0.9\textwidth]{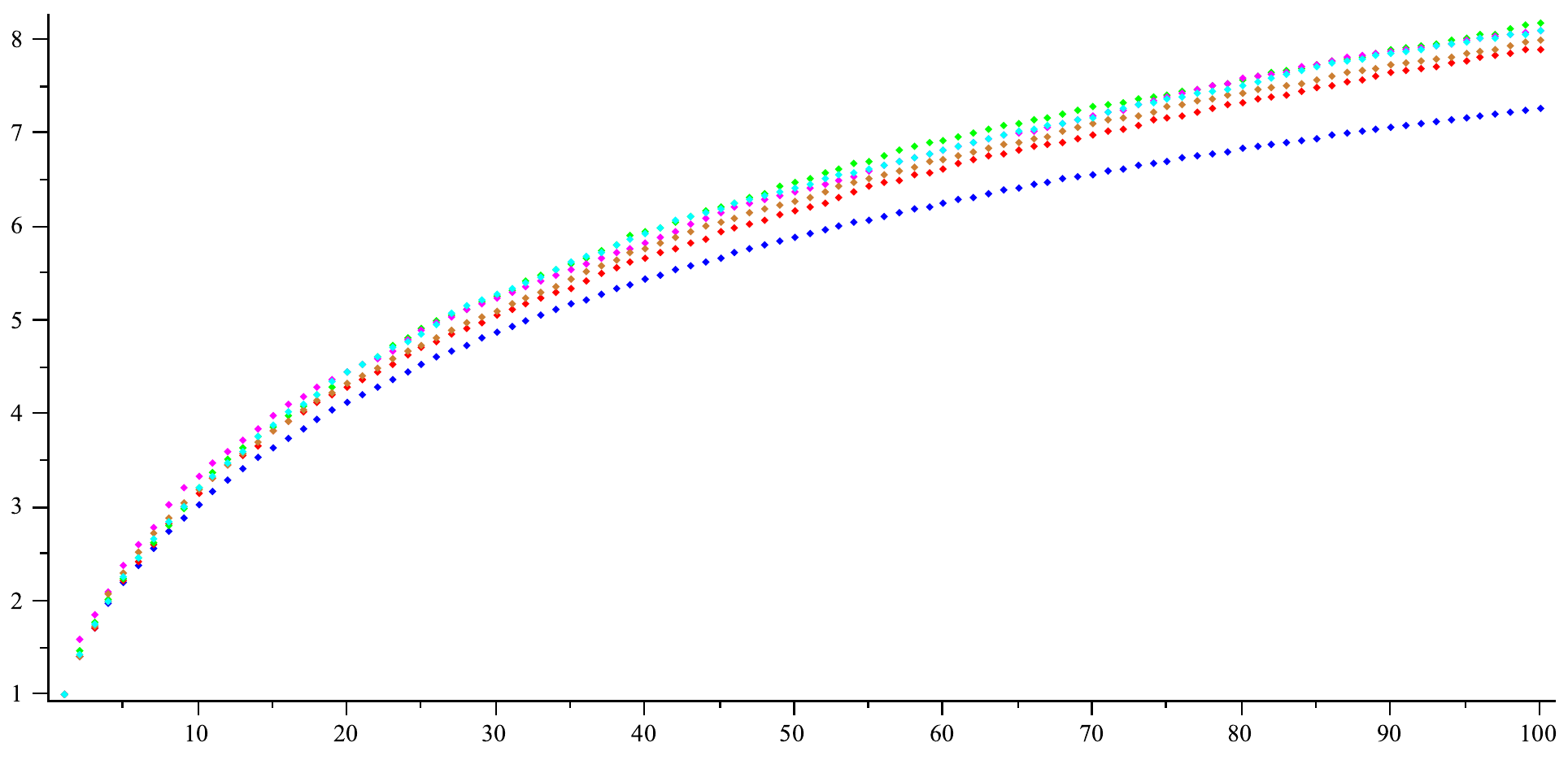}
\put(-425,110){$\frac{f_N}{\langle f_1\rangle}$}
\put(-200,-10){$\scriptstyle{N}$}\\[1em]
\caption{The enhancement $f_N/\langle f_1\rangle$ for $\langle\alpha_2\rangle = 0.58 > \frac{\pi}{2}$ (upper plot) and $\langle\alpha_2\rangle = 0.56 < \frac{\pi}{2}$ (lower plot).
Each plot for $\sigma(f_1) = 0$ (blue dots) and five realizations of a uniform distribution with $\sigma(f_1) = 0.35$.\label{fig-random-poly-2}}
\end{figure}

We are now ready to discuss the enhancement on a vertex of a completely general polytope, where the $N$ facets intersecting at the vertex can have different distances $f_1^{(i)}$ from the origin and different dihedral angles $\alpha_2^{(ij)}$ (with $i,j=1,\ldots, N$ and $i < j$). Analogous to what we did above for the simplified case with a single angle $\alpha_2$, the enhancement can again be derived iteratively by means of a recurrence relation for $f_n$. Since the expressions are quite lengthy, we refrain from explicitly writing them down here and only discuss the results of our analysis. A detailed explanation of the iteration procedure is given in Appendix \ref{gen-poly}. Let us only note here that the computation is in principle straightforward and can easily be done also for large $N$ using computer algebra.

In principle, our algorithm allows to compute the enhancement of the axion decay constant for any multi-axion model with arbitrary form of the scalar potential. We leave a detailed study of the phenomenology of such models for future work and only work out some generic characteristic features here.

In Fig. \ref{fig-random-poly-1}, we have plotted the enhancement $f_N/f_1$ for different $N$ while keeping $\langle \alpha_2 \rangle$ fixed. Here,
\begin{equation}
\langle \alpha_2 \rangle = \frac{2}{N(N-1)} \sum_{j} \sum_{i<j} \alpha_2^{(ij)}
\end{equation}
denotes the expectation value of $\alpha_2^{(ij)}$ on the vertex in question. Note that this is in general different from the expectation value of \emph{all} dihedral angles in the polytope.
In the plot, we considered different realizations of uniformly distributed $\alpha_2^{(ij)}$ with standard deviation
\begin{equation}
\sigma(\alpha_2) = \sqrt{\langle \alpha_2\vphantom{}^2 \rangle - \langle \alpha_2 \rangle^2}.
\end{equation}
An interesting observation is that variations in the angles tend to increase the enhancement at large $N$ compared to the case $\sigma=0$ of the previous section. More precisely, let us define $P\big(\frac{f_N}{\langle f_1\rangle}/\big(\frac{f_N}{\langle f_1\rangle}\big)_{\sigma=0} > 1\big)$ as the empirical probability that the enhancement for given $N$ and $\langle\alpha_2\rangle$ is larger for $\sigma \neq 0$ than the corresponding enhancement for $\sigma=0$. We have explicitly verified that $P\big(\frac{f_N}{\langle f_1\rangle}/\big(\frac{f_N}{\langle f_1\rangle}\big)_{\sigma=0} > 1\big) \to 1$ as $\sigma$ is increased by testing our recurrence relation for different distributions of $\alpha_2^{(ij)}$ (see Table \ref{table-prob}).\footnote{We stress again that this should not be confused with the landscape statistics of \cite{Bachlechner:2014gfa}. Here, we are not considering any string theory input on the likely shape of the polytope. We merely test our general recurrence relation for different parameter choices and then present the results in terms of statistical quantities.}

Another interesting observation is that the qualitative behavior of the enhancement is well-described already by the simple $\sigma=0$ model: as one increases $N$ while keeping $\langle \alpha_2 \rangle$ fixed, one passes through different regimes of enhancement with different scaling laws until one reaches an upper bound for which the enhancement diverges. This can again be understood geometrically: when the dihedral angles become large enough, the different facets intersecting at the vertex become almost parallel and the distance between the vertex and the origin becomes infinite.
Furthermore, the observation that we scan through the different enhancement regimes when $N$ is increased and $\langle \alpha_2 \rangle$ is kept fixed also implies that the separation of the regimes in terms of $\langle \alpha_2 \rangle$ must go to zero for large $N$. In other words, the larger $N$ becomes, the less we have to adjust $\langle \alpha_2\rangle$ in order to move from one regime to another. This is again analogous to the $\sigma=0$ case, where we found that the different regimes are separated by shifts in $\alpha_2$ of order $\mathcal{O}(1/N)$.

\begin{table}[t]\renewcommand{\arraystretch}{1.2}\setlength{\tabcolsep}{7pt}
\begin{center}
  \begin{tabular}{ |c || c | c | c || c | c | c | }
    \hline    
    & \multicolumn{3}{|c||}{$\sigma(\alpha_2) \neq 0$} & \multicolumn{3}{|c|}{$\sigma(f_1) \neq 0$}  \\ \hline
    & $\sigma=0.007$ & $\sigma=0.029$ & $\sigma=0.058$ & $\sigma=0.029$ & $\sigma=0.35$ & $\sigma=0.52$  \\ \hline\hline
    $\langle \alpha_2\rangle = 1.58 $ & 0.52 & 0.88 & 0.98 & 0.60 & 0.68 & 1.00 \\ \hline
    $\langle \alpha_2\rangle = 1.56 $ & 0.50 & 0.86 & 1.00 & 0.84 & 1.00 & 1.00 \\ \hline
    \end{tabular}
\caption{$P\big(\frac{f_N}{\langle f_1\rangle}/\big(\frac{f_N}{\langle f_1\rangle}\big)_{\sigma=0} > 1\big)$ at $N=100$ for two different values of $\langle \alpha_2\rangle$, one larger than $\frac{\pi}{2}$ and one smaller, and for two different variation patterns, one with $\sigma(\alpha_2) \neq 0$ and one with $\sigma(f_1) \neq 0$. As $\sigma$ is increased, $P\big(\frac{f_N}{\langle f_1\rangle}/\big(\frac{f_N}{\langle f_1\rangle}\big)_{\sigma=0} > 1\big) \to 1$ in all cases. Each case was tested for $50$ realizations.}
\label{table-prob}
\end{center}
\end{table}

In addition to variations in the angles, we also investigated the effect of variations in the distances $f_1^{(i)}$ (see Fig. \ref{fig-random-poly-2}). Compared to angle variations, this effect turned out to be smaller even for relatively large $\sigma(f_1)$. However, we found again that variations in $f_1$ tend to increase the enhancement instead of decreasing it, in the statistical sense discussed above (see Table \ref{table-prob}).

\section{Quantum gravity constraints}
\label{constraints}

As we have discussed above, general bottom-up models with multiple axions allow a parametric enhancement of the effective axion decay constant for a wide range of the parameters. For $f_\text{eff} \lesssim M_\text{p}$, there is no reason to suspect that anything goes wrong with such an enhancement, while for larger $f_\text{eff}$ it is possible that string theory forbids a further enhancement due to the WGC \cite{ArkaniHamed:2006dz, Rudelius:2014wla, Rudelius:2015xta, Montero:2015ofa, Brown:2015iha} (or, more generally, due to other possible quantum gravity constraints). If this is true, we should only expect to obtain scalings like those found above for moderately large $N$, whereas the enhancement should start to converge to a finite value beyond a certain threshold number $N_0$. In this section, we will not assess the validity of the different arguments for and against a violation of a possible bound on the field excursion. Instead, we will try to answer the question how a convergence could be ensured in a given model descending from a string compactification under the assumption that the WGC forbids a super-Planckian field excursion for axions.

\subsection{General considerations}
\label{constraints-1}

Let us consider a string compactification with $N>N_0$ axions, where $N_0$ is a threshold beyond which the parametric enhancement of the effective axion decay constant is stopped by virtue of the WGC. The following possibilities then come to mind:
\begin{itemize}
 \item The maximally allowed value for the individual decay constants of each axion decreases for large $N$ in such a way that the naive scaling of the effective axion decay constant along some diagonal is cancelled.\footnote{If there are more than $N$ terms in the scalar potential, it is not clear how to assign a decay constant to each axion. As noted above, a more useful definition is then to denote by ``individual decay constants'' the distances of the facets of the $N$-polytope to the origin.}
 \item The renormalization of the 4D Planck mass becomes relevant and scales with $N$ at least as strongly as the naive enhancement.
 \item The number $N$ is bounded from above, i.e., string theory forbids the existence of compactifications in which the number of axions exceeds $N_0$.
 \item Additional instanton terms in the potential bound the fundamental domain to stop the enhancement when $N$ grows too large, i.e., $P \gg N$ for models with $N>N_0$.
 \item The bound on the field excursion is violated in models with multiple axions.
\end{itemize}
Let us discuss these possibilities in detail. The first option may not be compatible with the WGC, for the following reason. In the T-dual picture of U(1) gauge fields \cite{Brown:2015iha}, the WGC imposes not only a bound on the mass of electrically charged particles but also on magnetically charged ones, with magnetic charge $1/g$ \cite{ArkaniHamed:2006dz}. This suggests that, for axions, both $f\to\infty$ and $f \to 0$ lead to the appearance of objects with a small action that modify the EFT. In the limit $f\to0$, one would expect that, instead of an instanton, the action of a magnetic dual becomes small, such as the Euclidean strings discussed in \cite{Montero:2015ofa}. From the point of view of string theory, one furthermore observes that letting $f \to 0$ can lead to a loss of perturbative control. In string compactifications, the size of $f$ is often related to the volume of certain cycles of the internal manifold (see, e.g., the single-axion examples discussed in \cite{Banks:2003sx}). If one makes $f$ too large, these cycles become small such that large string corrections would invalidate the EFT description. On the other hand, if one makes $f$ very small, the cycles become too large and KK modes become light, again leading to a break-down of the EFT.
It is interesting to note, however, that the individual decay constants in certain type IIB compactifications exhibit an $N$-dependence at fixed cycle volumes such that they shrink if $N$ is made large \cite{Grimm:2007hs}. Hence, these models seem to evade the simple arguments made above and precisely realize the mechanism of option one. In view of the above discussion, it would be important to understand how general this scaling behavior is and whether/how it is affected by the magnetic weak gravity conjecture. We defer a more thorough analysis of these open questions to future work and will assume in the following that there is no fundamental obstruction to constructing models with multiple axions in which the individual decay constants are slightly but not parametrically smaller than $M_\text{p}$.

The second option does not seem to solve the problem either since general arguments involving black holes suggest that the renormalized Planck mass grows at most like $\sqrt{N}$ \cite{Dvali:2007hz}. Moreover, the effect of renormalization is actually less severe when evaluated in concrete string models \cite{Bachlechner:2014gfa} (at least for those corrections that are explicitly known). Hence, the renormalization factor is not large enough to stop the power-law and exponential enhancements observed in \cite{Bachlechner:2014gfa, Choi:2014rja}.

Option three is hard to exclude in full generality. However, one may argue that, even if there was an upper bound on the number of axions allowed in string compactifications, this would unlikely be the mechanism responsible for stopping the parametric growth of $f_\text{eff}$. After all, it is well-known that, e.g., compactifications on Calabi-Yau 3-folds can easily yield several hundred or even thousands of axions such that the hypothetical bound $N_0$ would have to lie above that number.
This would imply a huge enhancement several orders above the Planck mass. We should therefore expect that another mechanism is responsible for stopping the enhancement way before a bound on $N$ could cut it off. A possible caveat is that moduli stabilization issues might lead to problems already for smaller $N$. Such problems were discussed in \cite{Blumenhagen:2014nba, Hebecker:2014kva} in the somewhat different context of axion monodromy inflation but are not directly applicable to models without monodromies (see also \cite{Dong:2010in, Buchmuller:2015oma} for possible effects of heavy moduli on the inflationary potential). In any case, the main issue discussed in these works seems to be tuning rather than a fundamental obstruction. This suggests that the worst-case scenario is that inflationary models with many axions are rare in the string landscape but not forbidden.

This leaves us with the possibility that additional (multi-)instanton terms become relevant in the scalar potential whenever the naive field range is too large. The appearance of unsuppressed multi-instanton corrections was indeed the reason why engineering super-Planckian decay constants failed in the single-axion models of \cite{Banks:2003sx}, and it was recently also emphasized as the main obstacle in the context of multi-axion models \cite{Montero:2015ofa, Brown:2015iha}. It is therefore natural to investigate the effect of such instanton corrections in a general model with multiple axions. In particular, it should be interesting to know \emph{how many} dominant instanton corrections would actually have to appear in the scalar potential in order to forbid a parametric enhancement of the effective axion decay constant.

A natural guess is that higher harmonics of D-brane instantons (i.e., multi-wrapped D-branes) may become unsuppressed at large $N$ and thus bound the field range. However, in order for this mechanism to be effective, multi-instantons with a huge charge would have to become relevant in the scalar potential. For example, in the simplest case of $N$-flation, cutting off the diagonals with $f_\text{eff}\sim\sqrt{N}$ would require the multi-instantons with charges $(\pm 1,\pm 1,\pm 1,\ldots,\pm 1)$ under the $N$ axions to have an action of order $1$. Hence, the actions of the leading instantons (i.e., those with unit charge) would have to be tiny, $S_E \sim 1/\sqrt{N}$. In a general model with power-law or exponential enhancement, the actions would have to be even smaller. Since the action of a D-brane scales with the volume of the wrapped cycle and inversely with the dilaton, this is expected to be the case only at extremely small volumes or at a huge string coupling, and thus we would be far away from the perturbative regime of string theory. In a perturbative string compactification with $f^{(i)}\lesssim M_\text{p}$, on the other hand, the instanton actions should be much larger. Indeed, gravitational (multi-)instantons, which are conjectured to be the low-energy effective descriptions of D-brane (multi-)instantons \cite{Montero:2015ofa}, have been shown in this regime to be highly suppressed at large $N$ such that they would not spoil enhanced field excursions in a potential generated by other instantons \cite{Montero:2015ofa, Bachlechner:2015qja}.

It is in principle conceivable that string theory simply does not admit any perturbatively controlled compactifications with many axions.
However, it is a priori not clear why this should be the case (cf. bullet point 3). Moreover, there exist proposed counter-examples in which higher harmonics of the known instanton effects are subdominant and super-Planckian field excursions appear to be possible \cite{Bachlechner:2014gfa, Hebecker:2015rya}.\footnote{The field excursion in the model of \cite{Bachlechner:2014gfa} is only marginally super-Planckian with $\Delta\phi\approx 1.13 M_\text{p}$. It would be important to see whether similar models with larger field ranges can be constructed. It has furthermore been pointed out that the model might be subject to dangerous corrections potentially spoiling the large field excursion \cite{Brown:2015iha}. A possible worry about the model of \cite{Hebecker:2015rya} is that it is a string-inspired scenario rather than a fully explicit string theory computation.} Assuming that the approximations made in these constructions are trustworthy, one may then wonder what could possibly happen in such a model in order to evade the conclusion that super-Planckian field excursions are allowed in string theory. If the WGC implies a bound on the field excursion, it would then predict additional non-perturbative effects that would have to become dominant in the scalar potential.

It is not the goal of this paper to perform explicit string theory computations to shed light on this important issue. Instead, we are interested in drawing model-independent conclusions from an analysis of the geometric properties of the fundamental domain of the axion moduli space, analogous to what we did in Section \ref{feff}. Such a field theory analysis can obviously not decide whether the WGC holds in any of its forms. Proving or disproving this would require input from quantum gravity (e.g., via an  analysis of instantons in string theory), but a field theory computation may still be useful in determining the general conditions under which the field excursion can be bounded. Motivated by the above discussion, we will specifically be interested in the question how many dominant instantons would have to be provided if the WGC indeed bounds the field excursion.

In the next section, we will focus on precisely this situation, i.e., we assume a model with $N$ axions where the individual decay constants $f^{(i)}$ are sub-Planckian but not parametrically smaller than $M_\text{p}$. We then ask how many dominant instantons would have to contribute to the scalar potential in such a model to 
ensure that there is no parametrically enhanced diagonal anywhere in field space. We will find evidence that this number grows exponentially with $N$ such that string theory would have to provide an enormous number of extra instanton corrections on top of the known non-perturbative effects, which are of order $N$ in typical string compactifications in regimes of perturbative control.

To summarize, it is not clear whether the alternatives discussed in this section suffice to explain a bound on the field range in string compactifications with many axions. This may suggest the existence of a loophole such that super-Planckian field excursions are actually allowed in string theory when the number of axions is large.

We should nevertheless stress that we have certainly not excluded any of the possibilities discussed in this section but merely pointed out several difficulties which are not explained by what is currently known in the literature. In order to understand each of these mechanisms in more detail, it is clearly important to analyze explicit string compactifications and check whether some of the above discussed problems are evaded. Such an analysis is beyond the scope of the present work.

\subsection{Relation between instantons and enhancement}

As discussed above, let us now consider a model with $N$ axions in which the individual decay constants are smaller but not parametrically smaller than $M_\text{p}$. We would now like to understand how many dominant instantons $P$ have to contribute to the scalar potential in order that the effective axion decay constant is not parametrically enhanced with $N$.\footnote{Our computation does not depend on the physical origin of the instanton corrections such that it would also apply to a situation where higher harmonics provide some of the $P$ dominant contributions to the scalar potential. However, as stated above, evidence in the literature suggests that these are highly suppressed for the known instantons in the regime of interest.}

So far, we have only studied ``local'' properties of the $N$-polytope bounding the axion moduli space without any input from string theory. In particular, we analyzed in Section \ref{feff} how the enhancement in the direction of a particular vertex depends on the distances and dihedral angles of the $N$ facets intersecting at that vertex. Now we would like to understand the ``global'' problem of how the enhancement is related to the total number of facets $2P$ of the polytope, which is directly related to the number of dominant instanton corrections in the scalar potential.

What result should we expect? As a simple example, consider a model in which $P=N$ and there is no alignment. The polytope is then an $N$-cube with Pythagorean enhancement $f_\text{eff} \sim \sqrt{N}$. How many extra instanton terms would have to appear in the scalar potential in order to evade a parametric growth of the enhancement at large $N$? A natural guess is that one would at least have to cap off each of the $2^N$ vertices of the cube by adding a further ``diagonal'' facet.
Hence, if there is a bound on the axion field excursion, the true fundamental domain consistent with quantum gravity would have to be a polytope with $P \sim 2^N$ instead of $P=N$. This would imply that string theory must generate an exponentially large number of dominant instantons in such a model. In the following, we will argue that a similar conclusion actually holds for \emph{generic} polytopes.

To see this, we will proceed in two steps. First, we will combine known results in the math literature with some general considerations to relate the distribution of the angles $\alpha_2^{(ij)}$ to $P$. We will then use our general recurrence relation to determine the convergence radius of $f_N$ in terms of $P$, i.e., we will analyze how large $P$ has to be chosen in order that $f_N$ converges to a finite value at large $N$. We will not attempt to give an analytic proof of the convergence radius since this is extremely difficult without specifying a particular polytope. Instead, we will use our recurrence relation to test the enhancement for several classes of generic polytopes and see whether it converges.

For simplicity, let us restrict to the case where the distances of all facets of the polytope from the origin are equal, i.e., $f^{(i)}_1 = f_1$, and only variations in the dihedral angles are allowed. This is physically reasonable as we typically expect the $f_1^{(i)}$ to be of roughly the same size. Furthermore, we have shown that introducing variations in the distances increases the enhancement such that our results below should be considered as providing a lower bound on the number of required instantons.

There are several results in the math literature which will be useful for our purpose. First, under the above assumption, we can map our problem of interest to a well-known statistics problem, namely the properties of randomly chosen points on the unit $N$-sphere. Taking these points to be the endpoints of the $2P$ normal vectors defining our polytope, their distribution is all we need to determine the dihedral angles and, via our recurrence relation, the enhancement of the effective axion decay constant.
We choose a uniform distribution of points on the hypersphere, which should correspond to the least amount of alignment, again giving a lower estimate for the expected enhancement of a general polytope.
It was proven in \cite{Cai:2013} that, for large $P$, the probability density for the angles between the normal vectors is then given by
\begin{equation}
p(\Theta)=\frac{\Gamma\left(\frac{N}{2}\right)}{\sqrt{\pi}\Gamma\left(\frac{N-1}{2}\right)} (\sin \Theta)^{N-2}. \label{x}
\end{equation}
Using that the total number of angles between the $2P$ normal vectors is $\left(\begin{smallmatrix}2P\\ 2\end{smallmatrix}\right)$, one can furthermore show \cite{Cai:2013} that
\begin{equation}
\Theta_\text{min} \to \arccos \sqrt{1-\e^{-4\frac{\ln P}{N}}}, \qquad \Theta_\text{max} \to \pi - \arccos \sqrt{1-\e^{-4\frac{\ln P}{N}}}
\end{equation}
in probability, where $\Theta_\text{min}$ and $\Theta_\text{max}$ are the smallest and the largest angles between any two normal vectors in the polytope (i.e., those angles for which the probability drops to $\frac{1}{P(2P-1)}$). For polynomially growing $P$, this implies
\begin{equation}
\Theta_\text{min} \to \frac{\pi}{2} - 2 \sqrt{\frac{\ln P}{N}}, \qquad \Theta_\text{max} \to \frac{\pi}{2} + 2 \sqrt{\frac{\ln P}{N}}.
\end{equation}
Hence, for polynomially growing $P$, all angles are likely to be very close to $\frac{\pi}{2}$, whereas, for exponentially growing $P$, the difference $\Theta_\text{max} - \Theta_\text{min}$ converges to a finite value and thus also allows smaller or larger angles.
For large $N$ and small $\left|\frac{\pi}{2}-\Theta\right|$, one can furthermore show \cite{Cai:2013} that \eqref{x} converges to
\begin{equation}
p(\Theta) = \frac{\sqrt{N-2}}{\sqrt{2\pi}} \exp \left[ -\frac{1}{2} \left( \sqrt{N-2} \left(\frac{\pi}{2}-\Theta\right) \right)^2 \right]. \label{normal}
\end{equation}
Hence, for polynomially growing $P$, the angles are normally distributed.

\begin{figure}[t]
\centering
\includegraphics[trim = 0mm 80mm 0mm 30mm, width=1\textwidth]{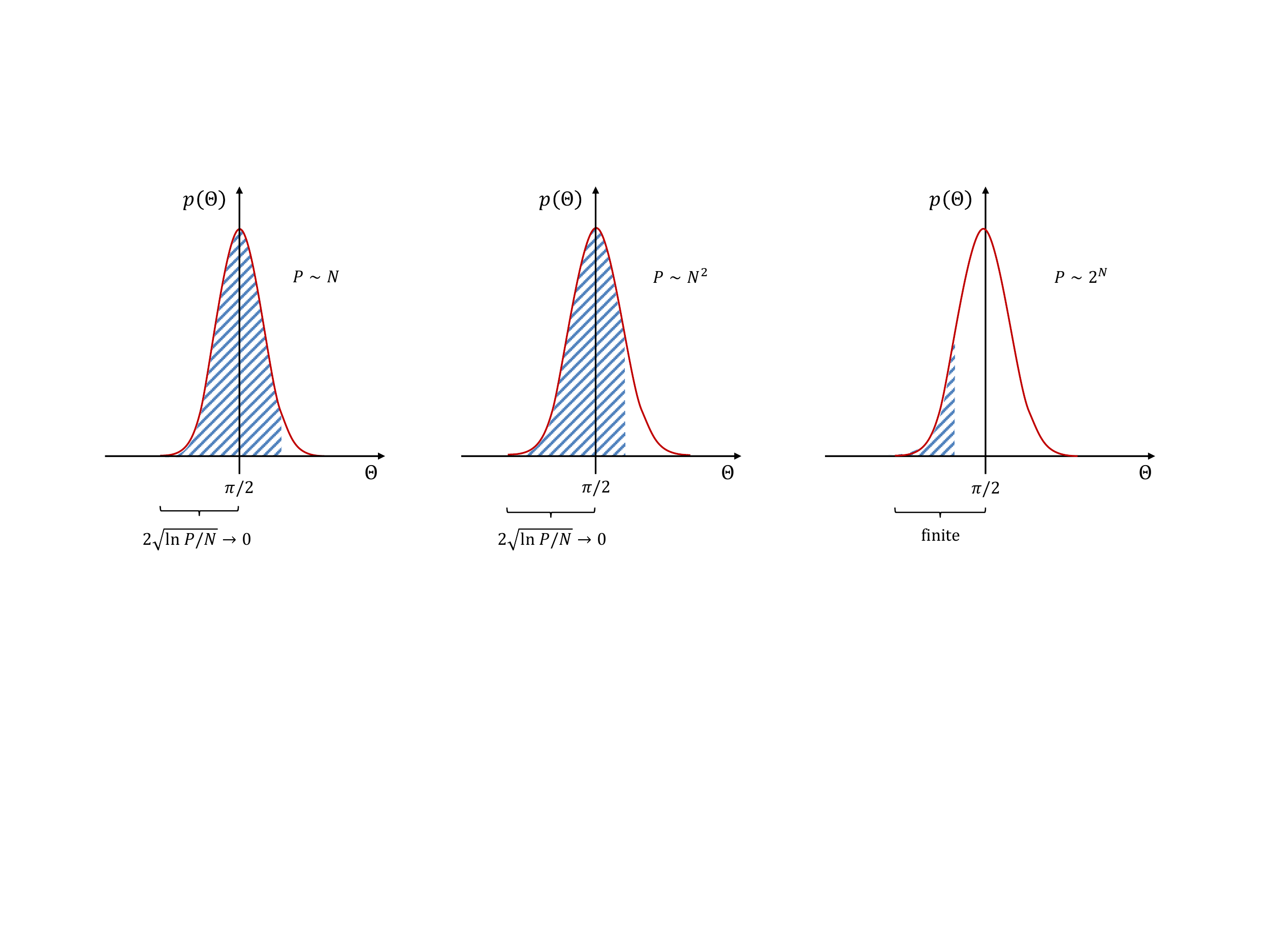}
\caption{Schematic plot of the angle distribution at large $N$ for two polytopes with polynomially growing $P$ (left and middle picture) and one with exponentially growing $P$ (right picture). The deviation of the minimal and maximal angles from $\frac{\pi}{2}$ goes to zero at large $N$ unless $P$ grows at least exponentially. The ratio of dihedral angles (shaded area) to the total number of angles typically gets smaller as $P/N$ gets larger.\label{fig-distributions}}
\end{figure}

\begin{figure}[p]
\centering
\includegraphics[width=0.9\textwidth]{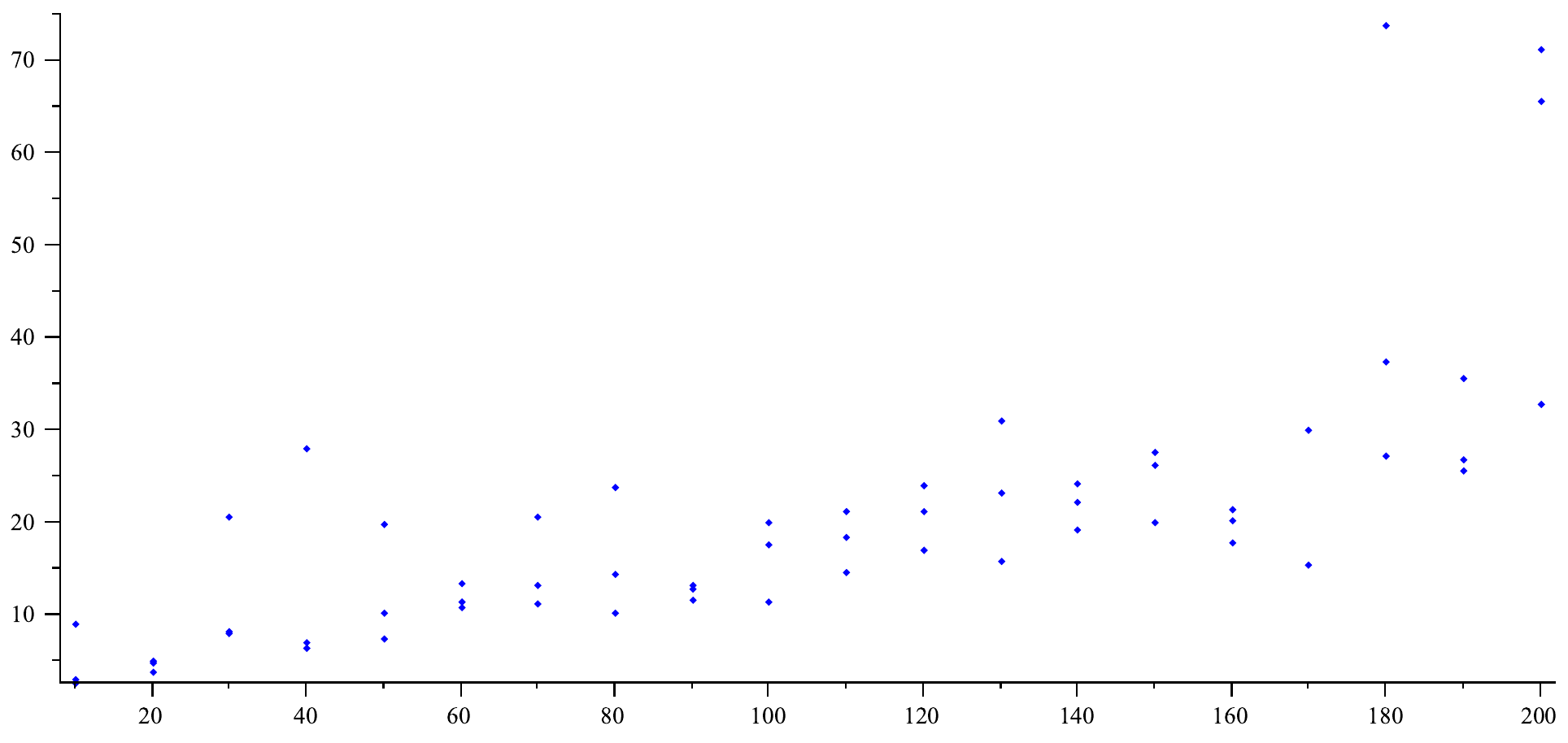}
\put(-430,100){$\frac{f_N}{f_1}$}
\put(-200,-10){$N$}\\[1em]
\includegraphics[width=0.9\textwidth]{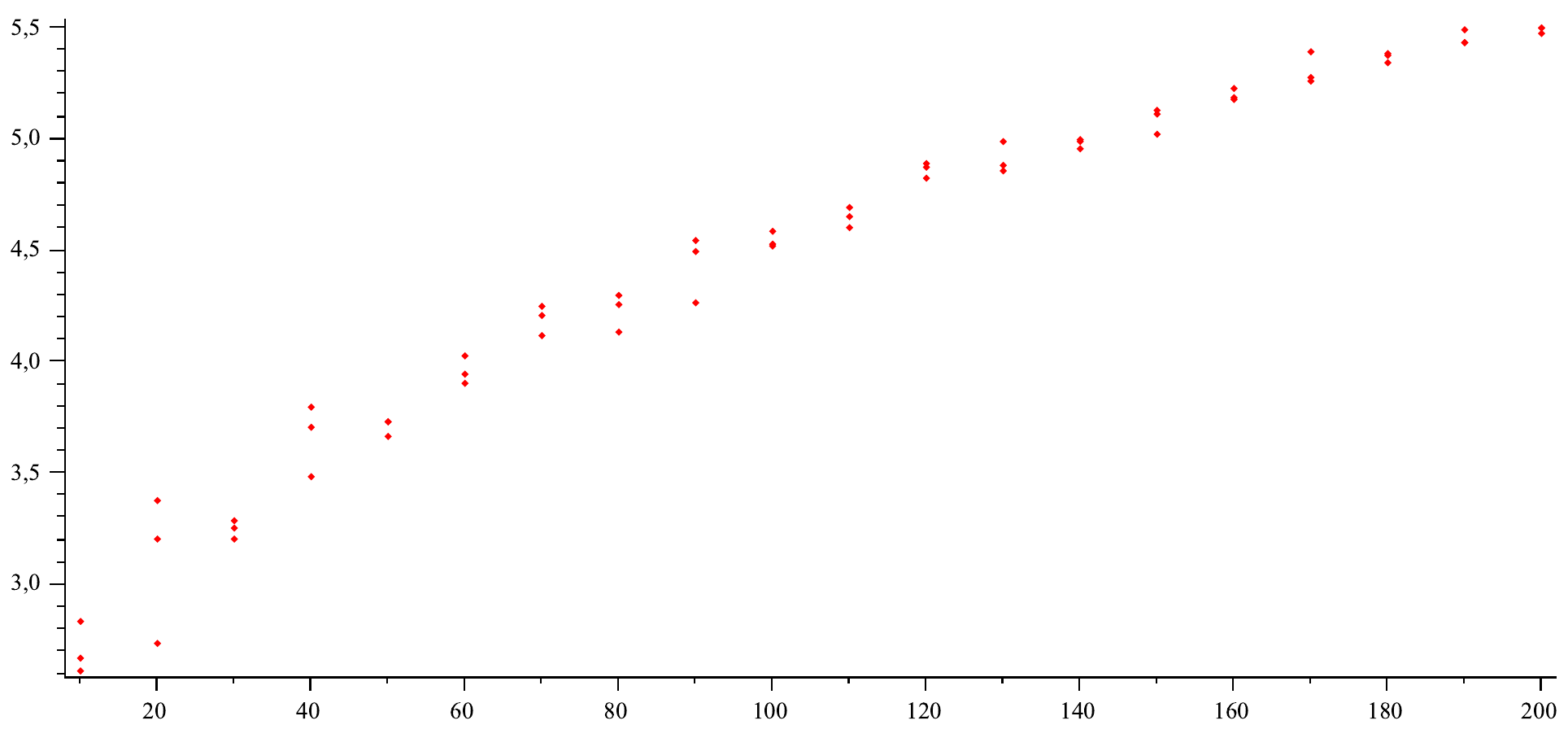}
\put(-430,95){$\frac{f_N}{f_1}$}
\put(-200,-10){$N$}\\[1em]
\caption{An estimate for the enhancement on typical vertices of polytopes with $P\sim N$ for different $N$, where the codimension 2 face number equals that of an $N$-cube (blue) or saturates the lower bound theorem (red).
\label{fig-conv1}}
\end{figure}

\begin{figure}[p]
\centering
\includegraphics[width=0.9\textwidth]{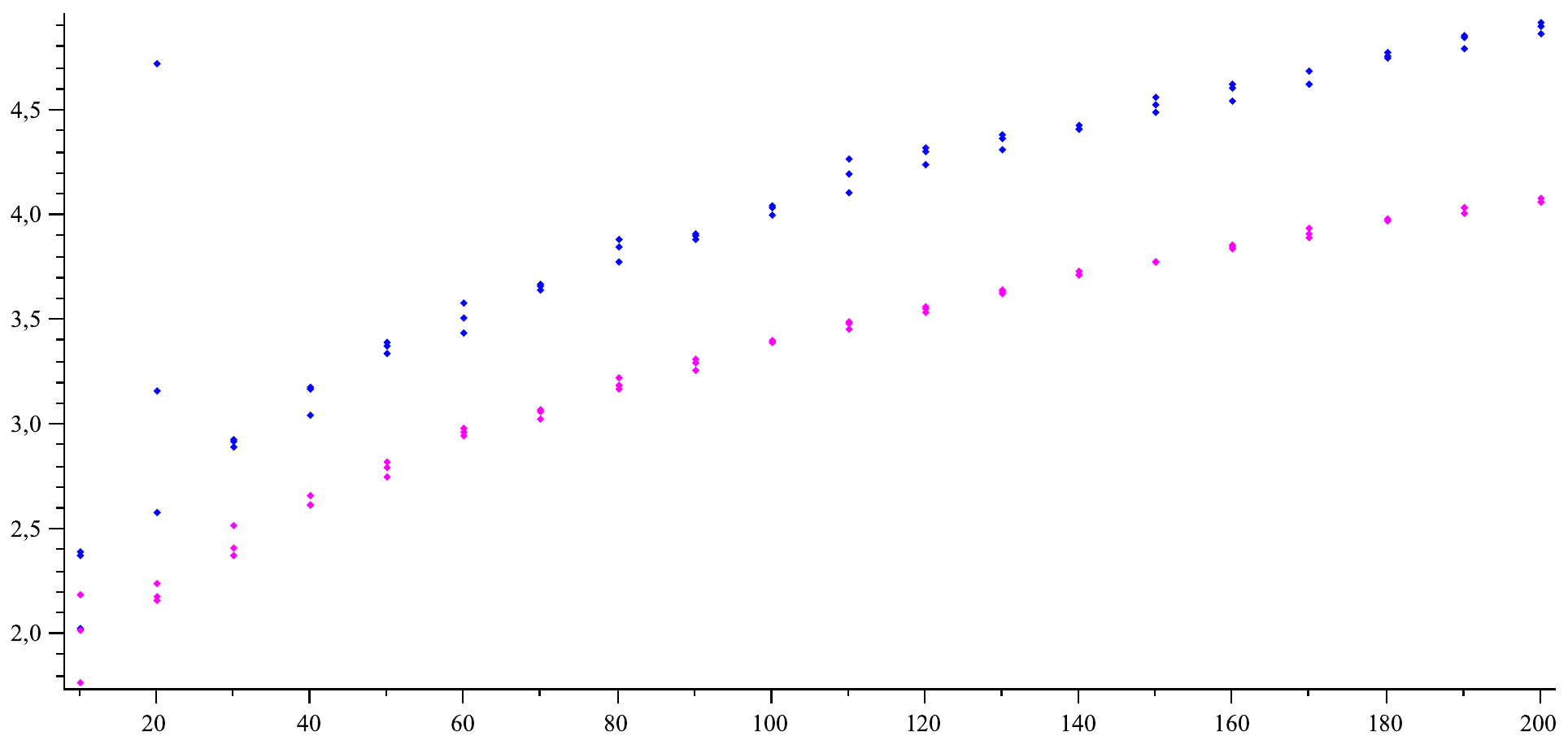}
\put(-430,100){$\frac{f_N}{f_1}$}
\put(-200,-10){$N$}\\[1em]
\includegraphics[width=0.9\textwidth]{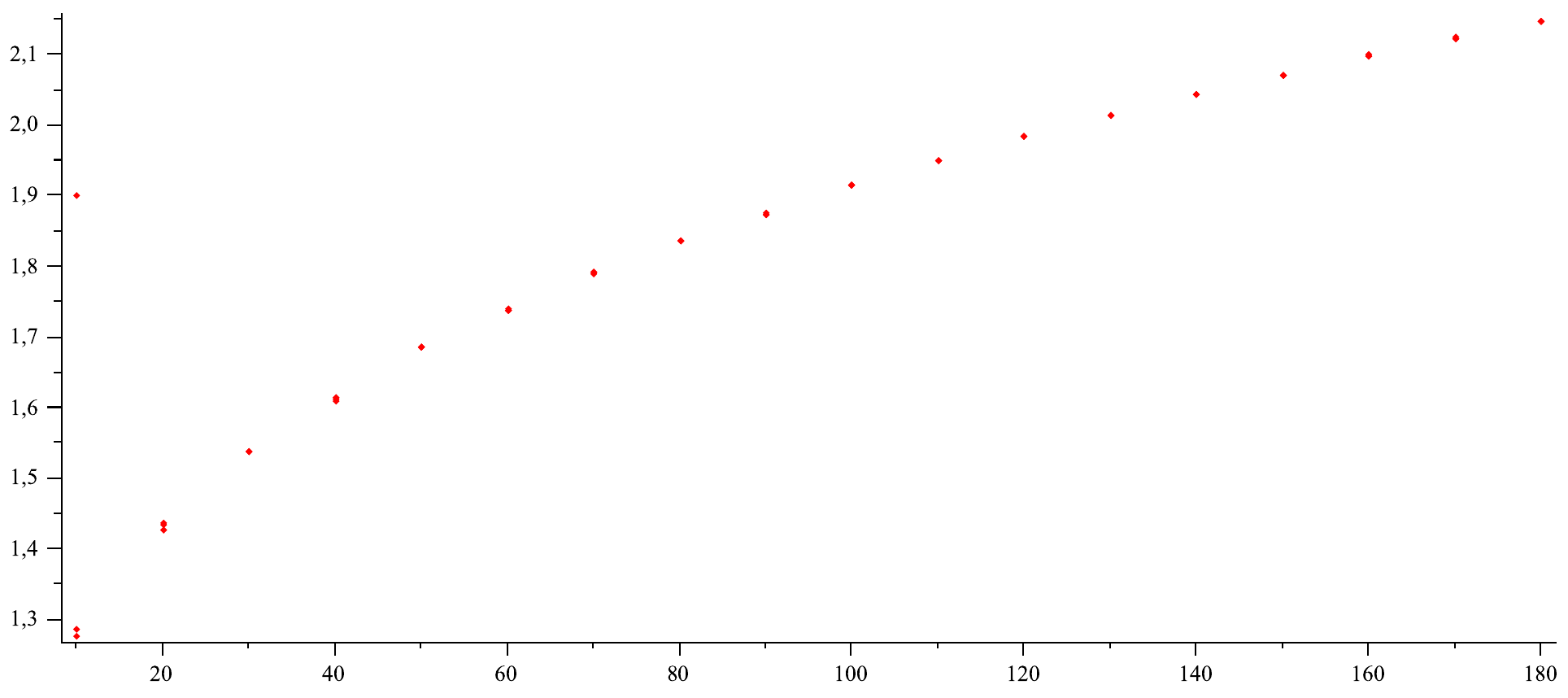}
\put(-430,95){$\frac{f_N}{f_1}$}
\put(-200,-10){$N$}\\[1em]
\caption{An estimate for the enhancement on typical vertices of polytopes with $P\sim N^2$ for different $N$, where the codimension 2 face number equals that of an associahedron (blue) or a cyclohedron (violet) or saturates the lower bound theorem (red).
\label{fig-conv2}}
\end{figure}

\begin{figure}[t]
\centering
\includegraphics[width=0.9\textwidth]{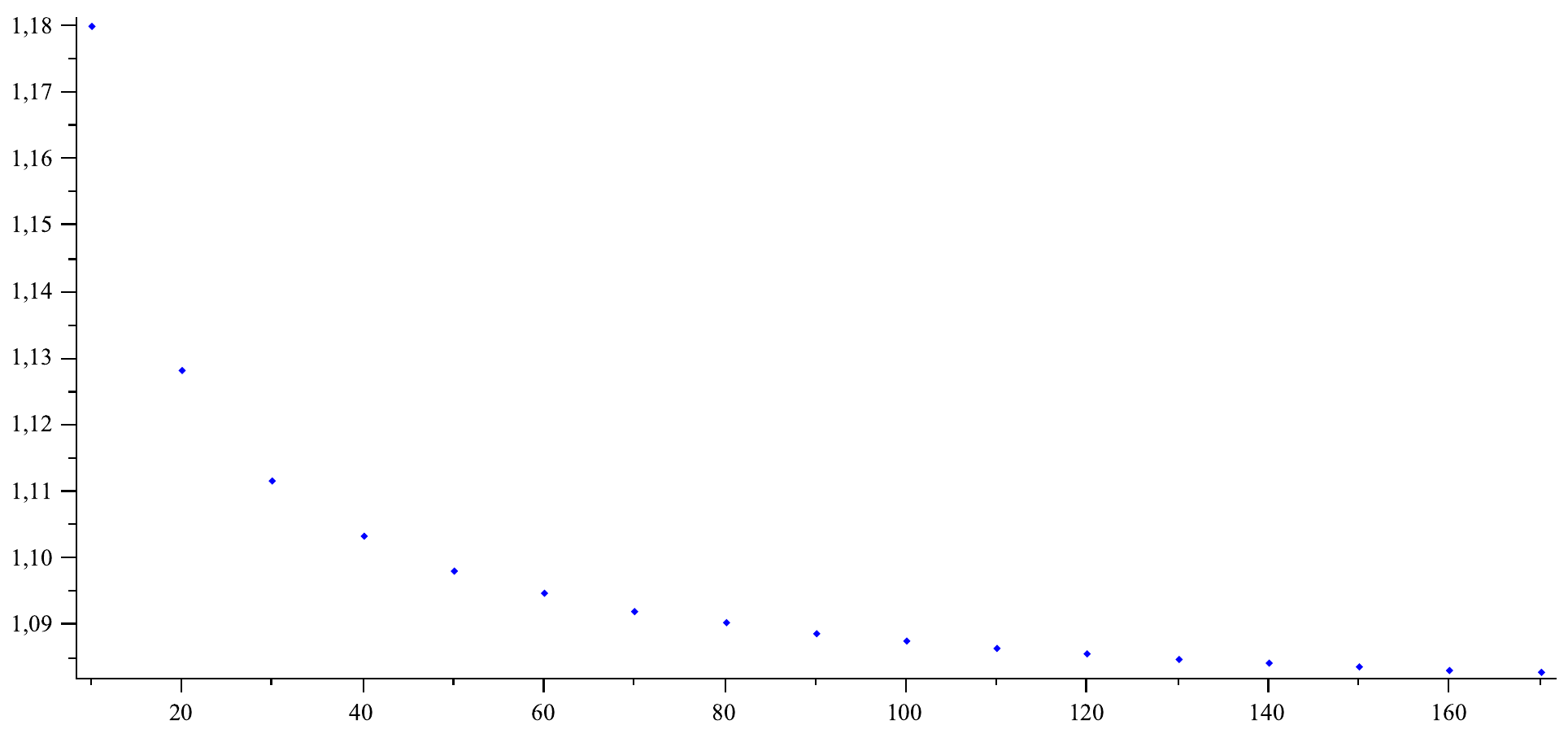}
\put(-430,100){$\frac{f_N}{f_1}$}
\put(-200,-10){$N$}\\[1em]
\caption{An estimate for the enhancement on typical vertices of a polytope with $P\sim 2^N$ for different $N$, where the codimension 2 face number equals that of a cross-polytope.
\label{fig-conv3}}
\end{figure}

For large $N$, the above results determine the distribution of the angles between any pair of normal vectors in the polytope. But what about the distribution of dihedral angles, i.e., the angles between those normal vectors whose facets intersect? If the number of facets is very small, they can effectively bend around the $N$-sphere such that every facet can intersect with every other facet in the polytope. In general, however, this will not be the case. The number of dihedral angles is therefore in general smaller than the total number of angles, and we should expect this ratio to become smaller as $P/N$ is increased. In an $N$-simplex, for example, the number of facets is very small, $2P = N+1$, and one can convince oneself that the number of dihedral angles equals the total number of angles $\left(\begin{smallmatrix}N+1\\ 2\end{smallmatrix}\right)$. In an $N$-cube, on the other hand, one has $2P=2N$. The total number of angles is then $\left(\begin{smallmatrix}2N\\ 2\end{smallmatrix}\right)$, where $N$ of them equal $180^\circ$ and the rest equals $90^\circ$. The number of dihedral angles is thus $\left(\begin{smallmatrix}2N\\ 2\end{smallmatrix}\right)-N$.

In general, the number of dihedral angles in a polytope heavily depends on its topology. It is therefore not possible to come up with a general formula unless we specify exactly the polytope we consider. Since the number of dihedral angles is given by the number of intersections of $(N-1)$-facets, it equals the number of $(N-2)$-facets, which we will denote by $F_{N-2}$ in the following. Hence, we can get a (conservative) estimate for the dihedral angles at a given vertex by drawing them from the $F_{N-2}$ smallest angles among the total $\left(\begin{smallmatrix}2P\\ 2\end{smallmatrix}\right)$ angles in the polytope, which in turn are drawn from the normal distribution \eqref{normal} (see Fig. \ref{fig-distributions}). This implies that, for a given $N$, the smaller the ratio is between $F_{N-2}$ and $\left(\begin{smallmatrix}2P\\ 2\end{smallmatrix}\right)$, the smaller are the dihedral angles and, hence, the enhancement.

One way to determine $F_{N-2}$ is to consider specific examples for which the face numbers are known and then study random polytopes under the assumption that their $F_{N-2}$ is the same. These could either be deformations of the known examples with the same topology or, more generally, other polytopes which happen to have the same $F_{N-2}$.
Two simple classes of examples are the $N$-cube with $2P=2N$ and $F_{N-2}= 2N(N-1)$ and the cross-polytope with $2P=2^N$ and $F_{N-2}=\frac{1}{2}N 2^N$. In addition, we consider two classes of polytopes with $P\sim N^2$, namely the so-called associahedra and cyclohedra (see, e.g., \cite{Goodman:2004, Miller:2007} for their exact definitions and properties). One finds that cyclohedra have $2P=N(N+1)$ and $F_{N-2} =  \frac{1}{4}(N+2)(N+1)N(N-1)$, while associahedra have $2P=\frac{1}{2}N(N+3)$ and $F_{N-2}=\frac{1}{12}(N+4)(N+3)N(N-1)$.\footnote{These numbers can be determined from so-called $f$-vectors or $h$-vectors, which encode the topology of polytopes. An explicit expression for the $f$-vector of a cyclohedron can be found, e.g., in \cite{Morton:2007}, and the $h$-vector of the polytope dual to an associahedron is given in \cite{Miller:2007}.}

A more general approach is to make use of a lower bound on $F_{N-2}$, which is available for a large class of polytopes called simple polytopes. These are polytopes whose vertices have degree $N$, i.e., $N$ facets intersect at each vertex. In fact, this is not a strong restriction for our purpose since we are only interested in the enhancement. In particular, any polytope can be made into a simple polytope by truncating all vertices with degree higher than $N$ (i.e., cutting these vertices off). If the truncation is done infinitesimally far away from a vertex, the enhancement in this direction does not change. Hence, our computations should be valid for any kind of polytope.
A useful theorem is then the so-called lower bound theorem, which was proven in \cite{Barnette:1971, Barnette:1973} and implies
\begin{equation}
F_{N-2} \ge 2NP - \frac{N(N+1)}{2} \label{lbt}
\end{equation}
for all simple polytopes.
The downside of this method is that the above bound is not very strong in the sense that it often massively underestimates the number of dihedral angles (and, hence, the enhancement). Taking as an example the associahedra and cyclohedra mentioned above, the actual number of dihedral angles scales like $N^4$, while the lower-bound theorem only gives a scaling $N^3$.

Following the strategy outlined above, we have plotted the enhancement along a typical vertex of several classes of example polytopes for different $N$, where ``typical'' refers to the statistical approach explained above. We considered random deformations of the cube ($P=N$) and the cross-polytope ($P\sim 2^N$) as well as two classes of random polytopes with $P\sim N^2$ whose codimension 1 and 2 face numbers (i.e., $2P$ and $F_{N-2}$) were assumed to equal those of associahedra or cyclohedra (see Figs. \ref{fig-conv1}--\ref{fig-conv3}). Note that we did not have to explicitly construct all these polytopes. Rather, we used our recurrence relation to compute the enhancement along a single vertex whose dihedral angles are determined using the algorithm described above.

Not surprisingly, one observes that, for the case $P=N$, the enhancement grows parametrically at large $N$. Since almost all angles are dihedral angles for an $N$-cube at large $N$, it is likely that many of them are larger than $\frac{\pi}{2}$. As was shown in Section \ref{feff}, we then expect a wide range of possible enhancements, which is indeed reproduced by the upper plot in Fig. \ref{fig-conv1}.\footnote{In fact, several realizations showed an infinite amount of enhancement because the angles were so large that the different facets could not ``close'' into a convex object anymore. This is consistent with Section \ref{feff}, where we found that small deviations from $\frac{\pi}{2}$ of order $\mathcal{O}(1/N)$ can already make the enhancement diverge, and with \cite{Choi:2014rja}, where it was argued that a moderate tuning of the anomaly coefficients can already lead to an exponentially large enhancement at large $N$.}
More interestingly, we can see from Fig. \ref{fig-conv2} that even a quadratic growth $P\sim N^2$ is not sufficient to stop the parametric enhancement. Although the large number of facets has the effect of making most of the dihedral angles smaller than $\frac{\pi}{2}$, this is apparently not enough to stop the enhancement. In Section \ref{feff}, we argued that deviations $\left|\frac{\pi}{2}-\langle\alpha_2\rangle\right| \sim \mathcal{O}(1/N)$ can already be sufficient to move away from the Pythagorean regime to a regime of slower enhancement, but we did not specify a precise criterion for when the enhancement stops entirely. Fig. \ref{fig-conv2} shows that having as much as $P \sim N^2$ facets is not enough. On the other hand, we can see from Fig. \ref{fig-conv3} that, as expected, an exponential growth of $P$ has the effect that the enhancement dies out completely. In fact, the number of facets then grows so fast that the enhancement even falls for large $N$.

In addition to these examples, we made use of the lower bound theorem to also get an estimate of the enhancement in completely general simple polytopes (lower plots in Figs. \ref{fig-conv1} and \ref{fig-conv2}). The plots show that the enhancement then grows rather slowly. As explained above, this is not surprising as \eqref{lbt} bounds the expected enhancement from below in a rather crude way. The interesting point to notice is, however, that, even under our very conservative assumptions, there is still no sign of convergence. This provides further evidence that not even a quadratic growth of $P$ with respect to $N$ is sufficient to stop a parametric enhancement of the axion field range.

Intuitively, these results can be understood from the interplay of two effects. First, when $P$ is made large while keeping $N$ fixed, the ratio of the number of dihedral angles to the total number of angles tends to get smaller such that the values of the dihedral angles move to the left-hand side of the normal distribution (see Fig. \ref{fig-distributions}). Hence, large $P$ tends to make the enhancement smaller for a given $N$. Second, at large $N$, almost all angles gather around the value $\frac{\pi}{2}$, where the width of the distribution is infinitesimally small unless $P$ grows at least exponentially with $N$. Hence, for polynomially growing $P$, the dihedral angles cannot deviate much from $\frac{\pi}{2}$ such that only an exponential growth of $P$ is effective in bounding the enhancement. This is of course not a proof but it justifies the expectation that an enormous number of instanton corrections would be required in order to restrict the axion field range to sub-Planckian values.

One might argue that the slow growth of the axion field range observed in the examples with $P\sim N^2$ is nothing to worry about. However, for sufficiently large $N$, this would still lead to a violation of any bound on the field excursion.
We should also again emphasize that our estimate for the enhancement was very conservative and in many cases heavily underestimates the $P$ necessary to stop the enhancement. As stated above, one reason is that we used the lower bound \eqref{lbt} on the number of dihedral angles for some of the plots, which is general but not very strong. Another reason is our assumption that the dihedral angles are always the smallest angles in the distribution of all angles, which need not be true in general. Furthermore, we computed the enhancement for a typical vertex, which we specified by drawing $\left(\begin{smallmatrix}N\\ 2\end{smallmatrix}\right)$ dihedral angles out of the set of all $F_{N-2}$ dihedral angles of the polytopes in question. However, a vertex determined this way is not necessarily the vertex with the largest enhancement in a given polytope. Finally, we did not take into account additional enhancement due to variations in the distances $f_1^{(i)}$. Hence, the above results should be seen as a very conservative lower estimate of the enhancement expected in a general polytope. Even in this restrictive situation, we found that a quadratic relation between $P$ and $N$ is not sufficient.

Another point is that, in string theory, one does not have the freedom to choose a polytope with the ``optimal'' relation between the number of facets and the enhancement produced. Hence, even if a class of polytopes existed at large $N$ for which the enhancement already converges with, say, $P \sim N^2$ (which is still a huge number), it would not be guaranteed that the axion moduli space in actual string theory models would have a fundamental domain in the shape of such a polytope.
A typical situation in a string theory based model is that a certain number of instanton corrections, say $N$, are already known to exist and be dominant. If these corrections are the leading instantons for the modes to which they couple and by themselves yield a fundamental domain in the shape of an $N$-cube (which is not far-fetched), it is difficult to escape the conclusion that the number of additional instantons required to cut off the Pythagorean enhancement would have to grow exponentially (cf. the discussion at the beginning of this section).

This is all not very surprising from the geometric point of view but it is interesting from the physics perspective.
Let us emphasize here that $P$ is not equivalent to the total number of (multi-)instanton corrections to the scalar potential but to the number of \emph{dominant} instanton corrections (i.e., those instantons which lead to non-negligible contributions to the scalar potential and thus bound the field range). Of course, the total number of multi-instantons can be infinite even in a model with only one axion but this is not what we claim here.
Furthermore, each of the $P$ instantons must couple to a different linear combination of the $N$ axions in order to create an additional facet in the polytope bounding the fundamental domain.
If one considers the alternatives discussed in Section \ref{constraints-1} unreasonable, our result thus suggests two different scenarios: either models with multiple axions are capable of exploiting a loophole that allows them to have super-Planckian field excursions or they must involve an enormous number of dominant instanton terms in the scalar potential. If quantum gravity indeed bounds the axion field range to be sub-Planckian in models with many axions, one would then have to explain where all the dominant instanton corrections come from.

\section{Discussion}
\label{conclusions}

In this paper, we studied the maximally allowed field excursion in models of inflation with multiple axions. We first showed that general bottom-up models admit a wide range of different regimes of enhancement, depending on the dihedral angles and distances of the facets in the polytope which bounds the fundamental domain of the axion moduli space. We also argued that the values of these parameters are related to the amount of alignment and the number of instanton terms in the scalar potential. We then asked the question which subset of these models is compatible with quantum gravity. We argued that models with a large number of axions must either involve an enormous number of dominant instanton terms in the scalar potential or be capable of super-Planckian field excursions by virtue of a loophole.
Whether or not such a loophole exists has been heavily debated in the recent literature. The various arguments in favor of and against a strict bound on the axion field range
are based on general quantum gravity considerations \cite{ArkaniHamed:2006dz, Rudelius:2014wla, delaFuente:2014aca, Rudelius:2015xta, Brown:2015iha, Brown:2015lia} as well as on explicit studies of instanton solutions \cite{Montero:2015ofa, Bachlechner:2015qja} and string theory models \cite{Rudelius:2014wla, Rudelius:2015xta, Bachlechner:2014gfa, Bachlechner:2015qja, Hebecker:2015rya}.

In this paper, we took a different approach to the problem.
Our aim was to draw general, model-independent conclusions from the hypothesis that the WGC forbids a parametric enhancement of the axion field range, based on a purely geometric reasoning. We found that, in order for the enhancement to converge to a finite value at large $N$, the number of dominant instanton terms in the scalar potential would have to grow faster than quadratically, presumably even exponentially, with $N$. In the perturbative regime of string theory, this is in stark contrast to the usual assumption of a linear relation between the number of instanton terms and the number of axions.
To our knowledge, explicit string compactifications show no evidence that such an enormous number of dominant instanton corrections should be expected. For example, the model of \cite{Denef:2005mm}, which was recently reconsidered in \cite{Bachlechner:2014gfa} in the context of axions, has $N=51$ and $P=60$. A possible caveat is that works on magnetized instantons \cite{Bianchi:2011qh} show that the number of non-perturbative corrections to the scalar potential is often larger than expected from a ``naive'' counting of the rigid divisors. Furthermore, it was emphasized in \cite{Montero:2015ofa} that also non-BPS instantons can contribute to the scalar potential in non-supersymmetric situations like inflation. Nevertheless, it is questionable whether these arguments could explain a number as huge as required by our findings. In particular, if the WGC bounds the field excursion, this is expected to be true also in supersymmetric setups even though they are not suitable for cosmology. Furthermore, it was argued in \cite{Bachlechner:2015qja} that the actions of the gravitational instantons of \cite{Montero:2015ofa} generically scale with powers of $N$ in large-$N$ models such that the corresponding corrections to the scalar potential tend to be suppressed.

To summarize, we anticipate that our result can be interpreted in two different ways. If one insists that the field excursion should be strictly bounded in models consistent with quantum gravity, it implies that bottom-up models of large-field inflation with multiple axions lie in the swampland since their assumptions on the number of terms in the scalar potential are false. Consistent models would instead have a much larger amount of dominant instanton corrections, which is only expected in non-perturbative regimes of string theory. On the other hand, assuming that large-$N$ models and alignment mechanisms can be embedded in perturbative regimes of string theory, one may argue that it is unreasonable to expect a number of dominant instantons as huge as required by the above results.
This suggests that a more reasonable interpretation of our results is to conclude that axions must be capable of violating bounds on the field excursion (imposed by the WGC or any other quantum gravity constraint), e.g., by exploiting one of the loopholes hypothesized in \cite{delaFuente:2014aca, Montero:2015ofa, Brown:2015iha}.

For future work, it would be interesting to further elaborate on the arguments put forward in this paper. In particular, it would be nice to get more precise results for the convergence radius of $f_\text{eff}$ in terms of $P$. It might also be interesting to revisit some of the alternatives discussed in Section \ref{constraints-1} in order to see whether axions can somehow evade our conclusions. Although we have given arguments for why we consider these alternatives unlikely, it would be useful to make these arguments watertight or else find possibilities to circumvent them. In view of our results, it would also be important to gain a better understanding of instanton solutions in general string compactifications with gravitational and gauge degrees of freedom (see also \cite{Brown:2015lia} for a discussion of this point) and to construct explicit string theory models to check our claims and those of \cite{Montero:2015ofa, Brown:2015iha} in more detail. We hope to report on further progress on these open questions in future work.

\section*{Acknowledgments}

I would like to thank Gary Shiu for getting me interested in this subject. I also thank Joe Conlon, Billy Cottrell, Thomas Grimm, Tiefeng Jiang, Anshuman Maharana, Luca Martucci, Christoph Mayrhofer, Gary Shiu, Pablo Soler and Fang Ye for useful discussions and comments. Finally, I am grateful to the HKUST Jockey Club Institute for Advanced Study for hospitality during a visit where part of this work was completed. This work was supported by the DFG Transregional Collaborative Research Centre TRR 33 ``The Dark Universe''.
\\

\appendix

\section{Enhancement for a general polytope}
\label{gen-poly}

In this section, we compute the enhancement in the direction of a vertex of a completely general polytope, where the $N$ facets intersecting at the vertex can have different distances $f_1^{(i)}$ from the origin and different dihedral angles $\alpha_2^{(ij)}$ (with $i,j=1,\ldots, N$ and $i < j$).

\subsection{Trigonometry}

\begin{figure}[t]
\centering
\includegraphics[trim = 0mm 60mm 0mm 30mm, width=1\textwidth]{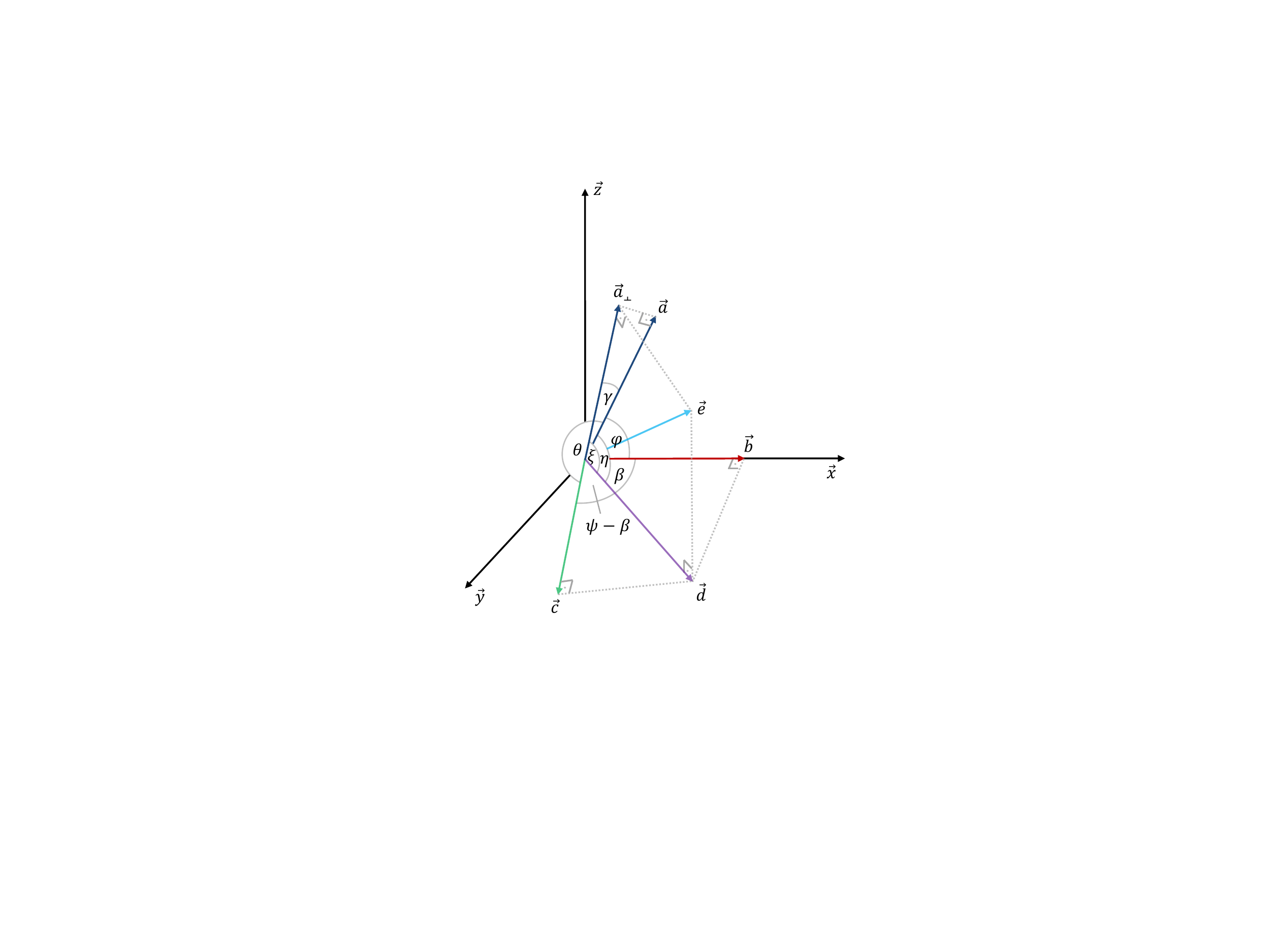}
\caption{Vectors and their mutual angles.
\label{fig-trig}}
\end{figure}

We first collect some trigonometric identities that will be useful below. The situation relevant for us is depicted in Fig. \ref{fig-trig}. Our goal is to compute the lengths of $\vec d$ and $\vec e$ depending on the lengths of $\vec a,\vec b,\vec c$ and on their mutual angles $\varphi,\psi,\theta$. All relations we will need below for the enhancement are then special cases of this situation.

Let us choose our coordinate system such that
\begin{equation}
\vec a = \begin{pmatrix} a_1 \\ a_2 \\ a_3 \end{pmatrix}, \quad\!\!\! \vec a_\perp = \begin{pmatrix} a_{\perp 1} \\ a_{\perp 2} \\ a_{\perp 3} \end{pmatrix}, \quad\!\!\! \vec b = b \begin{pmatrix} 1 \\ 0 \\ 0 \end{pmatrix}, \quad\!\!\! \vec c = c \begin{pmatrix} \cos \psi \\ \sin \psi \\ 0 \end{pmatrix}, \quad\!\!\! \vec d = d \begin{pmatrix} \cos \beta \\ \sin \beta \\ 0 \end{pmatrix}, \quad\!\!\! \vec e = \begin{pmatrix} e_1 \\ e_2 \\ e_3 \end{pmatrix}.
\end{equation}
From $\vec a \cdot \vec a = a^2$, $\vec a \cdot \vec b = ab \cos \varphi$ and $\vec a \cdot \vec c = ac \cos\theta$, we find
\begin{gather}
a_1 = a \cos \varphi, \quad a_2 = a \frac{ \cos \theta - \cos\psi \cos \varphi}{\sin \psi}, \\ a_3 = a \sqrt{1 - \frac{\cos^2 \varphi + \cos^2\theta - 2 \cos\psi \cos \varphi \cos\theta}{\sin^2 \psi}}.
\end{gather}
Furthermore, we have
\begin{equation}
\frac{b}{d} = \cos \beta, \quad \frac{c}{d} = \cos ( \psi - \beta),
\end{equation}
which can be solved for $d$ and $\beta$,
\begin{equation}
d = \frac{\sqrt{b^2 + c^2 - 2bc\cos \psi}}{\sin \psi}, \quad \beta = \arccos \frac{b \sin \psi}{\sqrt{b^2 + c^2 - 2bc\cos \psi}}.
\end{equation}

We now compute the angle $\gamma$ between $\vec a$ and the $(\vec z, \vec d)$ plane. In order to find it, we define a vector $\vec a_\perp$, which lies in the $(\vec z, \vec d)$ plane such that its angle with $\vec a$ is minimized. Demanding that $\vec a_\perp$ lies in the $(\vec z, \vec d)$ plane amounts to having
\begin{equation}
\vec z \times \vec a_\perp = \begin{pmatrix} -a_{\perp 2} \\ a_{\perp 1} \\ 0 \end{pmatrix} \propto \vec z \times \vec d = \begin{pmatrix} -d \sin \beta \\ d\cos \beta \\ 0 \end{pmatrix}.
\end{equation}
Hence, we can make the ansatz
\begin{equation}
\vec a_\perp = a_\perp \begin{pmatrix} \lambda \cos \beta \\ \lambda \sin\beta \\ \sqrt{1-\lambda^2} \end{pmatrix},
\end{equation}
where $a_\perp$ is the length of the vector and $\lambda$ is a yet unknown number. For later convenience, we choose the length $a_\perp$ such that $\vec a_\perp$ touches the intersection of the $(\vec z, \vec d)$ plane with the plane orthogonal to $\vec a$ (see Fig. \ref{fig-trig}),
\begin{equation}
a_\perp = \frac{a}{\cos\gamma}.
\end{equation}
The angle $\gamma$ between $\vec a$ and $\vec a_\perp$ is determined by
\begin{equation}
\vec a \cdot \vec a_\perp = a a_\perp \cos \gamma = a_\perp \left(a_1 \lambda \cos\beta + a_2\lambda\sin\beta + a_3 \sqrt{1-\lambda^2}\right),
\end{equation}
which yields
\begin{equation}
\gamma = \arccos \frac{a_1 \lambda \cos\beta + a_2\lambda\sin\beta + a_3 \sqrt{1-\lambda^2}}{a}.
\end{equation}
Minimizing with respect to $\lambda$, we find
\begin{equation}
\lambda = \frac{a_1\cos\beta + a_2 \sin\beta}{\sqrt{(a_1\cos\beta + a_2 \sin\beta)^2+a_3^2}}.
\end{equation}

The angle $\eta$ between $\vec a$ and $\vec d$ is given by
\begin{equation}
\vec a \cdot \vec d = a d \cos \eta = a_1 d \cos \beta + a_2 d \sin \beta,
\end{equation}
yielding
\begin{equation}
\eta = \arccos \frac{a_1 \cos \beta + a_2 \sin \beta}{a}.
\end{equation}
The angle $\xi$ between $\vec a_\perp$ and $\vec d$ is given by
\begin{equation}
\vec a_\perp \cdot \vec d = a_\perp d \cos \xi = a_\perp d \lambda.
\end{equation}
Hence,
\begin{equation}
\xi = \arccos \lambda.
\end{equation}
The vector $\vec e$ also lies in the $(\vec z, \vec d)$ plane. We only need its length, which is given by
\begin{equation}
e = \frac{\sqrt{a_\perp^2 + d^2 - 2a_\perp d \cos \xi}}{\sin \xi}.
\end{equation}
Finally, we will need to know the angle $\chi$ between $\vec a_\perp$ and a vector $\vec g$, which reaches into a fourth dimension,
\begin{equation}
\vec g = \begin{pmatrix} g_1 \\ g_2 \\ g_3 \\ g_4 \end{pmatrix},
\end{equation}
and whose components are specified in terms of its length $g$ and its mutual angles with $\vec a, \vec c, \vec d$,
\begin{equation}
\vec g^2 = g^2, \quad \vec a \cdot \vec g = ag \cos \delta, \quad \vec c \cdot \vec g = cg \cos \zeta, \quad \vec d \cdot \vec g = dg \cos \omega.
\end{equation}
Using these relations to compute the components of $\vec g$, we can substitute the latter into $\vec a_\perp \cdot \vec g = a_\perp g \cos \chi$ to find
\begin{align}
\chi = \arccos \Bigg[ & \lambda \cos\omega + \frac{\sqrt{1-\lambda^2}\,a\cos\delta}{a_3} \notag \\ & + \frac{\sqrt{1-\lambda^2}\left[ \cos\omega(a_1\sin\psi-a_2\cos\psi)-\cos\zeta(a_1\sin\beta-a_2\cos\beta)\right]}{a_3(\cos\psi\sin\beta-\cos\beta\sin\psi)} \Bigg].
\end{align}

\begin{figure}[t]
\centering
\includegraphics[trim = 0mm 60mm 0mm 10mm, width=1\textwidth]{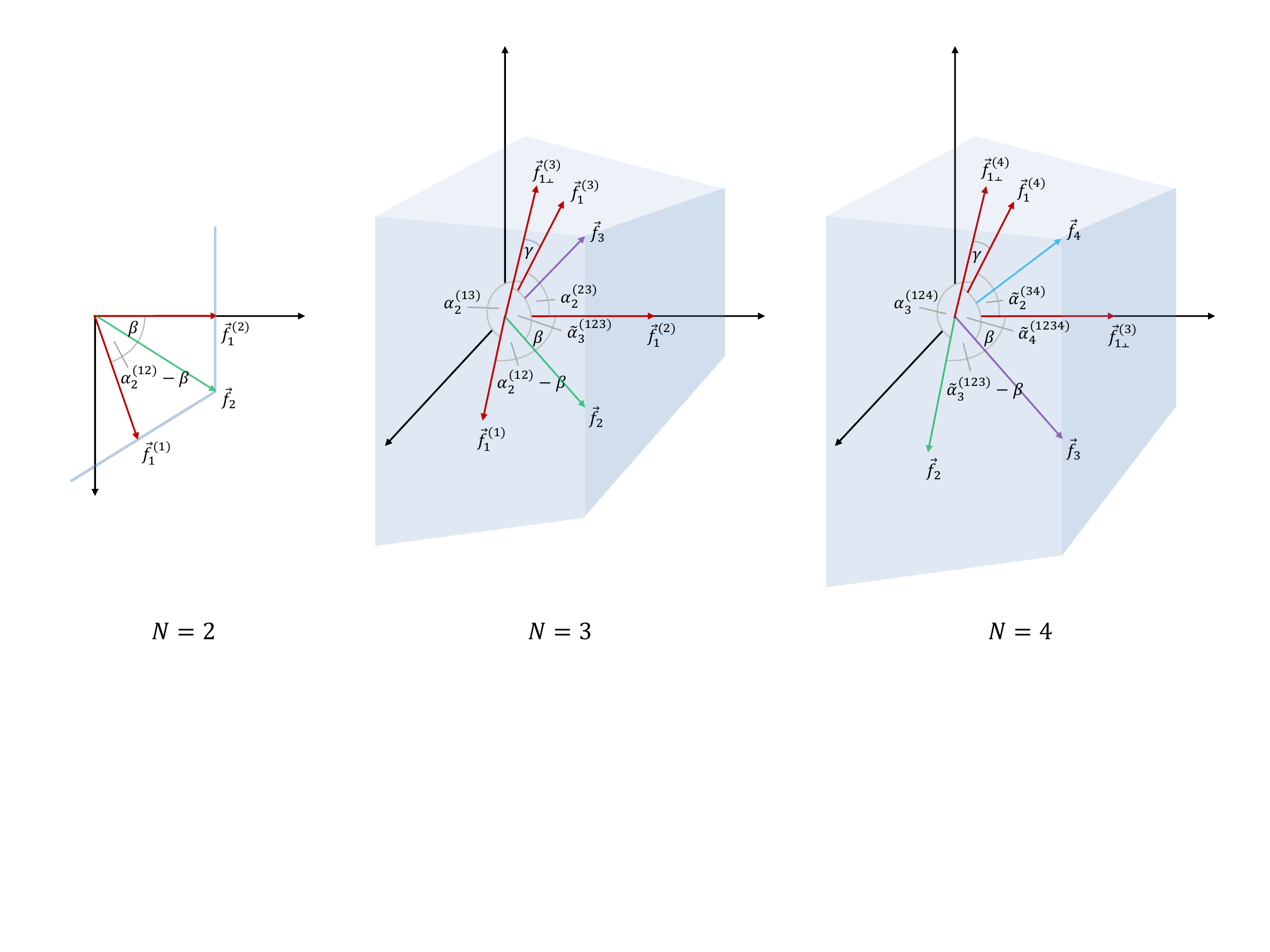}
\caption{Normal vectors and the angles between them in a general $N$-polytope for $N=2$, $N=3$ and $N=4$.
\label{fig-polytope-general}}
\end{figure}

\subsection{Recurrence relation}

Let us now compute the enhancement of the effective axion decay constant for $N=2$ and $N=3$. Hence, we consider a vertex defined by the intersection of three planes (see Fig. \ref{fig-polytope-general}). Let us denote the normal vectors to these planes by $\vec f_1^{(1)}, \vec f_1^{(2)}, \vec f_1^{(3)}$, their distances to the origin by $f_1^{(1)}, f_1^{(2)}, f_1^{(3)}$ and their dihedral angles by $\alpha_2^{(12)}, \alpha_2^{(13)}, \alpha_2^{(23)}$.  Furthermore, $\vec f_2$ denotes the vector which is normal to the intersection of the two planes that are parallel to the $\vec z$ axis, and $\vec f_3$ is the vector pointing to the vertex, with lengths $f_2$ and $f_3$, respectively. Finally, $\vec f_{1\perp}^{(3)}$ denotes the vector which is normal to the intersection of the $(\vec z, \vec f_2)$ plane with the plane normal to $\vec f_1^{(3)}$, and $\tilde\alpha_3^{(123)}$ is the angle between $\vec f_{1\perp}^{(3)}$ and $\vec f_2$. Our goal is to compute $f_2$ and $f_3$ in terms of $f_1^{(1)}, f_1^{(2)}, f_1^{(3)}$ and $\alpha_2^{(12)}, \alpha_2^{(13)}, \alpha_2^{(23)}$.

By comparing Fig. \ref{fig-polytope-general} and Fig. \ref{fig-trig}, one notes that this computation is identical to the one in the previous section upon identifying
\begin{equation}
\vec a = \vec f_1^{(3)}, \quad \vec a_\perp = \vec f_{1 \perp}^{(3)} \quad \vec b = \vec f_1^{(2)}, \quad \vec c = \vec f_1^{(1)}, \quad \vec d = \vec f_2, \quad \vec e = \vec f_3
\end{equation}
and
\begin{equation}
\psi = \alpha_2^{(12)}, \quad \varphi = \alpha_2^{(23)}, \quad \theta = \alpha_2^{(13)}, \quad \xi = \tilde \alpha_3^{(123)}.
\end{equation}
Hence, for $N=2$, we find
\begin{equation}
f_2 = \frac{\sqrt{ \big(f_1^{(1)}\big)^2 + \big(f_1^{(2)}\big)^2 - 2f_1^{(1)}f_1^{(2)}\cos \alpha_2^{(12)}}}{\sin \alpha_2^{(12)}}.
\end{equation}
For $N=3$, the expression is already rather lengthy,
\begin{equation}
f_3 = \frac{\sqrt{\Big(\frac{f_1^{(3)}}{\cos \gamma}\Big)^2 + f_2^2 - 2 \frac{f_1^{(3)}}{\cos \gamma} f_2 \cos \tilde \alpha_3^{(123)}}}{\sin \tilde \alpha_3^{(123)}}
\end{equation}
with
\begin{align}
& \tilde \alpha_3^{(123)} = \arccos \frac{a_1\cos\beta + a_2 \sin\beta}{\sqrt{(a_1\cos\beta + a_2 \sin\beta)^2+a_3^2}}, \\
& a_1 = f_1^{(3)} \cos  \alpha_2^{(23)}, \quad a_2 = f_1^{(3)} \frac{ \cos \alpha_2^{(13)} - \cos\alpha_2^{(12)} \cos \alpha_2^{(23)}}{\sin \alpha_2^{(12)}}, \\ & a_3 = f_1^{(3)} \sqrt{1 - \frac{\cos^2 \alpha_2^{(23)} + \cos^2\alpha_2^{(13)} - 2 \cos\alpha_2^{(12)} \cos \alpha_2^{(23)} \cos \alpha_2^{(13)}}{\sin^2 \alpha_2^{(12)}}}, \\
& \beta = \arccos \frac{f_1^{(2)} \sin \alpha_2^{(12)}}{\sqrt{\big(f_1^{(1)}\big)^2 + \big(f_1^{(2)}\big)^2 - 2f_1^{(1)} f_1^{(2)}\cos \alpha_2^{(12)}}}, \\ & \gamma = \arccos \Bigg[ \frac{\sqrt{(a_1 \cos\beta + a_2\sin\beta)^2+a_3^2}}{f_1^{(3)}} \Bigg].
\end{align}

For general $N$, the enhancement can be obtained iteratively by repeating the above computation for successive 3-dimensional subspaces of the full $N$-dimensional moduli space. This is illustrated in Fig. \ref{fig-polytope-general}.
Analogous to the notation used above, we now denote normal vectors to $(N-n)$-facets by $\vec f_n$ and their lengths by $f_n$. Furthermore, we define a vector $\vec f_{1\perp}^{(n)}$ in each 3-dimensional subspace by demanding that it lies in a plane orthogonal to $\vec f_{1\perp}^{(n-1)}$ and $\vec f_{n-2}$ for each $n\ge 4$ (see Fig. \ref{fig-polytope-general}). We also have to compute various angles between the different normal vectors, for which we choose the notation $\alpha_n^{(\ldots)} = \measuredangle(\vec f_1, \vec f_{n-1})$ and $\tilde \alpha_n^{(\ldots)} = \measuredangle(\vec f_{1\perp}, \vec f_{n-1})$. To avoid ambiguities, we indicate the subspace in which the angles are computed in superscript, e.g., $\alpha_3^{(124)}$ denotes an angle between normal vectors that lie in the 124 directions.

Each iteration $n=4,\ldots, N$ then involves three steps. First, comparing Fig. \ref{fig-polytope-general} and Fig. \ref{fig-trig}, we can identify
\begin{equation}
\vec a = \vec f_1^{(n)}, \quad \vec a_\perp = \vec f_{1\perp}^{(n)}, \quad \vec b = \vec f_{1 \perp}^{(n-1)}, \quad \vec c = \vec f_{n-2}, \quad \vec d = \vec f_{n-1}, \quad \vec e = \vec f_n \label{ident1}
\end{equation}
and
\begin{gather}
\psi = \tilde \alpha_{n-1}^{(123\ldots n-1)}, \quad \varphi = \tilde \alpha_2^{(n-1,n)}, \quad \theta = \alpha_{n-1}^{(123\ldots n-2,n)}, \quad \xi = \tilde \alpha_n^{(123\ldots n)}. \label{ident2}
\end{gather}
This gives us $f_{n}$ in terms of $f_{n-1}$ and our input parameters $f_1^{(i)},\alpha_2^{(ij)}$. Second, in order to obtain the angles $\alpha_n^{(123\ldots n-1,m)}$, which will be required in the next step, we have to consider
\begin{equation}
\vec a = \vec f_1^{(m)}, \quad \vec a_\perp = \vec f_{1\perp}^{(m)}, \quad \vec b = \vec f_{1 \perp}^{(n-1)}, \quad \vec c = \vec f_{n-2}
\end{equation}
and
\begin{gather}
\psi = \tilde \alpha_{n-1}^{(123\ldots n-1)}, \quad \varphi = \tilde \alpha_2^{(n-1,m)}, \quad \theta = \alpha_{n-1}^{(123\ldots n-2,m)}, \quad \xi = \tilde \alpha_n^{(123\ldots n-1,m)}, \quad \eta = \alpha_n^{(123\ldots n-1, m)}
\end{gather}
for $m=n+1,\ldots,N$. Finally, in order to obtain the angles $\tilde \alpha_2^{(n,m)}$ required for the next iteration, we have to consider, again for $m=n+1,\ldots,N$,
\begin{equation}
\vec g = \vec f_1^{(m)}, \quad \omega = \alpha_n^{(123\ldots n-1,m)}, \quad \zeta = \alpha_{n-1}^{(123\ldots n-2,m)}, \quad \delta = \alpha_2^{(n,m)}, \quad \chi = \tilde \alpha_2^{(n,m)},
\end{equation}
where $\vec a$, $\vec b$, etc. and $\psi$, $\varphi$, etc. are defined as in \eqref{ident1} and \eqref{ident2}.

Using this iteration rule together with the expressions computed in the previous section, we obtain a recurrence relation for $f_n$ in terms of $f_{n-1}$. It is then straightforward to automatize the iteration steps using computer algebra and determine the enhancement $f_N/\langle f_1\rangle$ for large $N$. The special case discussed in the main text is recovered by setting $f_1^{(i)}=f_1$, $\alpha_2^{(ij)}=\alpha_2$. This considerably simplifies the recurrence relation since it implies $\vec a_\perp = \vec a$ and, hence, $\gamma=0$ by symmetry.
\\

\bibliographystyle{utphys}
\bibliography{groups}

\end{document}